\documentclass[a4paper,11pt]{article}
\pdfoutput=1 

\usepackage{jheppub} 
\usepackage{slashed} 
\usepackage[T1]{fontenc} 
\usepackage[utf8]{inputenc}
\usepackage{subfig}

\def\hbbg{$H\rightarrow b\overline{b}g \;$}

\def\hbb{$H\rightarrow b\overline{b}\;$}
\def\hcc{$H\rightarrow c\overline{c}\;$}

\def\msbar{\overline{\text{MS}}}
\def\os{\text{OS}}
\def\eps{\epsilon}
\def\mmu2om2{\left (\frac{\mu^2}{m^2} \right )^{\!T\eps}}
\def\bb{{\bar b}}

\def\spa#1.#2{\langle #1 #2\rangle}
\def\spb#1.#2{[ #1 #2]}
\def\spab#1.#2.#3{\langle #1 |#2| #3] }

\newcommand{\nn}{\nonumber \\}
\def\top #1{\mathcal{T}_{#1}}

\newcommand{\GGvec}{\ensuremath \vec{\text{I}}}

\newcommand{\MM}{\ensuremath \mathbb{M}}
\def\dlog{d\log}
\newcommand{\GG}{\ensuremath \text{I}}

\def\amp{\mathcal{A}}
\def\order#1{\mathcal{O}\left(#1\right)}
\def\as{\alpha_s}

\title{ Top-induced  contributions to $H\rightarrow b\overline{b}$ and $H\rightarrow c\overline{c}$  at $\mathcal{O}(\alpha_s^3)$}

\author{Roberto Mondini,}         
\author{Ulrich Schubert,}       
\author{and Ciaran Williams}    

\affiliation{Department of Physics,\\University at Buffalo, The State University of New York, Buffalo
14260, USA}

\emailAdd{rmondini@buffalo.edu}
\emailAdd{ulrichsc@buffalo.edu}
\emailAdd{ciaranwi@buffalo.edu}

\begin{abstract}{
In this paper we present a fully-differential calculation for the contributions to the partial widths \hbb and \hcc that are sensitive to the top quark Yukawa coupling $y_t$ to order $\alpha_s^3$. 
These contributions first enter at order $\alpha_s^2$ through terms proportional to $y_t y_q$ ($q=b,c$). At order $\alpha_s^3$ corrections to the mixed terms are present as well as a new contribution proportional to $y_t^2$. Our results retain the mass of the final-state quarks throughout, while the top quark is integrated out resulting in an effective field theory (EFT). 
Our results are implemented into a Monte Carlo code allowing for the application of arbitrary final-state selection cuts. 
As an example we  present differential distributions for observables in the Higgs boson rest frame using the Durham jet clustering algorithm. We find that the total impact of the top-induced (i.e.~EFT) pieces
is sensitive to the nature of the final-state cuts, particularly $b$-tagging and $c$-tagging requirements. For bottom quarks, the EFT pieces contribute to the total width (and differential distributions) at around the percent level. The impact is much bigger for the \hcc channel, with effects as large as 15\%. We show however that their impact can be significantly reduced by the application of jet-tagging selection cuts. 
}
\end{abstract}

\begin{document} 
\maketitle
\flushbottom

\section{Introduction} 

The seminal moment in particle physics of the last twenty years was the discovery of the Higgs boson in 2012 by the ATLAS and CMS experiments at CERN's Large Hadron Collider (LHC)~\cite{Aad:2012tfa,Chatrchyan:2012xdj}. In the years since its discovery the continued study of the Higgs boson has become one of the key missions of high energy physics. The global properties of the Higgs boson 
(its spin, mass, and parity) are by now well constrained~\cite{Khachatryan:2014jba}. Going forward, of particular interest is the study of the Higgs boson couplings to other particles in the Standard Model (SM) and itself. The Higgs boson self-coupling, fully predicted in the SM from known parameters, awaits experimental verification (or contradiction) and represents a key to understanding the large-scale picture of Electroweak (EW) symmetry breaking.  

Over the course of Run I and Run II of the LHC, constraints on the couplings of the Higgs boson to particles in the SM have significantly improved~\cite{Khachatryan:2014jba,Sirunyan:2017exp,Aad:2015zhl,Aad:2015gba} and should continue to do so over the forthcoming Run III and subsequent HL-LHC runs over the next couple of decades~\cite{Cepeda:2019klc}. Complimentary to the continued study at a hadron 
machine, plans are afoot to construct a next-generation lepton collider~\cite{Gomez-Ceballos:2013zzn,Baer:2013cma,Abada:2019zxq}. Such a machine would offer a pristine environment 
in which to study the Higgs boson, with sub-percentage measurements of couplings across the board. 

Constraints on, and measurements of, the couplings of the Higgs boson proceed through measurements of its different production and decay mechanisms. Of the decay channels, \hbb is particularly important. For the 125 GeV Higgs boson, \hbb has a large branching fraction ($\sim$ 50\%) and therefore dominates the total decay width of the Higgs boson. Uncertainties in the measurement of the 
Higgs-bottom coupling propagate through to every measurement of the (on-shell) Higgs boson through its dependence on the total width. Extended Higgs sectors (which naturally arise in Beyond the Standard Model (BSM) scenarios) can alter the relative coupling of up-type and down-type quarks to the Higgs boson, so that constraining the \hbb process can in turn lead to constrains on extended Higgs sectors. For these reasons, it is thus highly desirable to study \hbb as accurately as possible. In addition, another long-term goal of the Higgs program is to directly measure the decay of the Higgs boson to the second-generation particles (specifically, muons and charms). Such a measurement would concretely establish the validity of the Higgs mechanism in the SM more broadly than existing third-generation and vector boson studies. A measurement of \hcc experimentally is extremely challenging, given the rampant QCD backgrounds, difficulties in charm-tagging, and smaller branching fraction. Nevertheless exciting progress has been made recently, with analyses reporting first direct constraints on $\sigma \times {\mathcal{BR}}$ in associated production at the LHC~\cite{Aaboud:2018fhh,Sirunyan:2019qia}.

Beyond the obvious challenges associated with making a sub-percentage precision measurement at the LHC or a Future Collider (FC), there are several theoretical issues which must be addressed. First and foremost, the parameter which is being extracted from the theory needs to carefully considered. At leading order in perturbation theory the identification $y_b= m_b/v$ is made, where $y_b$ is the bottom Yukawa coupling we seek to constrain, $m_b$ is the bottom-quark mass, and $v$ is the vacuum expectation value. At LO the partial width $\Gamma_{H\rightarrow b\overline{b}}$ is proportional to $y_b^2$, and thus $y_b$ could be readily extracted from the decay rate of \hbb (as part of a global fit).

The quantities in $y_b= m_b/v$  are defined at leading order in terms of bare Lagrangian parameters, and require appropriate renormalization at higher orders in perturbation theory. Although the bottom quark is not an isolated stable particle, for perturbative predictions of $\Gamma_{H\rightarrow b\overline{b}}$ it is treated as such, and accordingly it is convenient to use the on-shell renormalization scheme to define the mass at higher orders. Preserving the relationship $y_b= m_b/v$ would therefore also suggest evaluating the Yukawa coupling in the on-shell scheme. However, it was noted long ago in the first computations of \hbb at next-to-leading order~\cite{Braaten:1980yq} that employing the on-shell scheme results in large higher-order effects. These effects can be compensated if, instead of evaluating $y_b$ in the on-shell scheme, one chooses the $\msbar$ scheme. A feature of the $\msbar$ scheme is that it allows one to evolve the couplings to different scales, where for instance the bottom-quark mass entering the definition of the Yukawa coupling is evaluated not near the bottom pole mass, but at the Higgs boson mass. By capturing the relevant logarithms in the running the subsequent higher order predictions using $y_b^{\msbar}$ are more convergent. This then defines a mixed renormalization scheme, in which there are effectively two bottom quark masses: the $\msbar$ mass that enters the definition of the Yukawa coupling and the kinematic mass that enters the rest of the calculation (e.g. the propagators, completeness relations etc). When comparing to experimental data we therefore constrain $y_b^{\msbar}$. A further advantage of this scheme is that, by treating the kinematic bottom quark mass as different from the coupling mass, one can take a simplified limit where the former goes to zero (henceforth referred to as the ``massless'' approximation). 

Further complications arise at higher order (beyond the parameter definition just discussed) since we need to consider Feynman diagrams in which the bottom quark does not directly couple to the Higgs boson, but instead the Higgs couples to either top quarks or a massive vector boson, which then subsequently produce final state bottom quarks. Of particular interest in this paper is the expansion in QCD and the role of the top quark. At Next-to-Next-to Leading Order (NNLO) in QCD a term of the form $y_b y_t$ appears and Next-to-Next-to-Next-to Leading Order (N3LO) a term proportional to $y_t^2$ enters the prediction. Given the large hierarchy between $y_t$ and $y_b$ it is crucial to understand how large the new pieces are, and in particular how they manifest themselves in collider analyses. This requires precise knowledge of both inclusive, and differential predictions of \hbb at a suitably high order in perturbation theory. We note that in the massless limit the contribution from the $y_t y_b$ pieces vanishes, due to the presence of a helicity flip. These pieces are likely to be even more important in the decay \hcc due to the large hierarchy $y_t >> y_c$, and potential quasi-collinear enhancements. 

Much work has gone into computing inclusive rates for \hbb over several decades~\cite{Braaten:1980yq,Surguladze:1994gc,Larin:1995sq,Chetyrkin:1995pd,Chetyrkin:1996sr} and  as such, higher-order corrections from QCD for the inclusive decay width are known up to N$^4$LO (i.e.~up to order $\mathcal{O}(\alpha_s^4)$) \cite{Baikov:2005rw}. Additionally, the electroweak (EW) corrections have been known for some time~\cite{Dabelstein1992}, as well as the mixed QCD$\times$EW corrections ($\mathcal{O}(\alpha \alpha_s)$) \cite{Mihaila:2015lwa} and the two-loop master integrals for the mixed QCD$\times$EW corrections for the Higgs-top Yukawa coupling contributions to \hbb have also been computed~\cite{Chaubey:2019lum}. There has also been significant recent progress on computing differential predictions more relevant for collider analyses. Fully-differential predictions at NNLO in QCD were computed several years ago \cite{Anastasiou:2011qx,DelDuca:2015zqa} in the massless limit and more recently including the full kinematic effects of the bottom quark mass~\cite{Bernreuther:2018ynm,Behring:2019oci}. 
The massless limit was extended to a fully differential N3LO prediction recently~\cite{Mondini:2019gid,Mondini:2019vub}\footnote{A component of the N3LO calculation is the two-loop \hbbg amplitudes, which have been calculated in ref.~\cite{Ahmed:2014pka}.}.

In this paper we therefore aim to complete the computation at $\mathcal{O}(\alpha_s^3)$ by extending the results of ref.~\cite{Mondini:2019gid} to include the pieces proportional to $y_t$. In order to do so we will retain the mass of the bottom quark throughout the computation, but integrate out the top quark in order to produce an Effective Field Theory (EFT).  Our paper proceeds as follows, in Section~\ref{sec:calc} we outline the details of our calculation, including further discussion of the mixed renormalization scheme, and the IR structure of our calculation. Differential results for \hbb are presented in Section~\ref{sec:results} and for \hcc in Section~\ref{sec:hccres}. We draw our conclusions in Section~\ref{sec:conc}. Two appendices contain technical parameters and specifications related to our calculation. 

\section{Calculation}
\label{sec:calc}

\subsection{Overview} 
\begin{figure}
\begin{center}
\includegraphics[width=14cm]{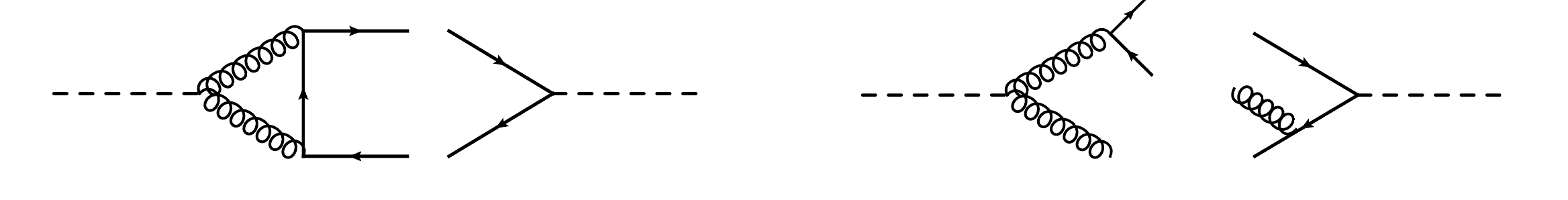}
\caption{Representative Feynman diagrams for the ``LO'' contributions to the processes discussed in this paper, there are two separate processes corresponding to two-body, and three-body decays of the Higgs. Both processes occur through an interference of a traditional $y_b$ amplitude, with an EFT induced amplitude.}
\label{fig:ytybas2}
\end{center}
\end{figure}

In this section we present an overview of our calculation, focusing for now on the decay \hbb. We begin by discussing the general features of the top-initiated contributions to \hbb. When performing this calculation, one of the initial decisions that must be made is how to handle the top quark itself. The complexity of the calculation can be reduced considerably if the top quark is integrated out of the theory, resulting in an effective field theory (EFT) of QCD with $n_l$ massless and $n_h$ massive flavors. In terms of the bare parameters (denoted by a $B$ label), the relevant parts of the subsequent Lagrangian can be written as
\begin{eqnarray}
\mathcal{L}&=&\mathcal{L}^{\rm{QCD}}_{g}
+\frac{1}{2}((\partial_{\mu}H)^2-m_H^2 H^2)-V_{SM}(H)
\nonumber\\&&+\sum_{j \in \{n_l\}} \overline{\psi}^B_j(i\slashed{D})\psi^B_j
\nonumber\\&&+\sum_{j \in \{n_h\}} \left(\overline{\psi}^B_j(i\slashed{D}-m^B_j)\psi^B_j
-y_j^B C_2^B \; H \overline{\psi}^B_j\psi^B_j \right) \nonumber\\&&
-\frac{C_1^B}{ v} H G_{\mu\nu}^{B,a}G^{\mu\nu}_{B,a}.
\label{eq:LHEFT}
\end{eqnarray}
In the above Lagrangian $C_1$ and $C_2$ are matching parameters which relate the EFT to the SM. The symbol $\slashed{D}$ defines the covariant derivative which couples the gluon field $A^B_{\mu}$ to the fermions $\psi^B_j$. The term $V_{SM}(H)$ defines the triple and quartic interactions of the Higgs boson (which are not needed for our calculation), and $\mathcal{L}^{\rm{QCD}}_{g}$ collects all of the kinetic terms for the gluons, ghost, and gauge-fixing terms required to define the QCD Lagrangian. 
We will discuss the relationship between the bare parameters and the renormalized ones in the next subsection. 

The differences arising between the full SM and the EFT have been studied for our process of interest at  $\mathcal{O}(\alpha_s^2)$ in ref.~\cite{Primo:2018zby}, finding very small differences between the two prescriptions, thus motivating the application of the EFT to $\mathcal{O}(\alpha_s^3)$. 
For the remainder of this paper we will work in the EFT defined above, but will frequently refer to the EFT pieces as ``top-induced'' for ease of discussion.

At $\mathcal{O}(\alpha_s^2)$ there are two distinct processes which give rise to top-induced contributions. These can be classified according to the number of final-state partons present (either two or three). Representative Feynman diagrams for the two different processes are presented in fig.~\ref{fig:ytybas2}, which correspond to two-body and three-body decays of the Higgs boson.  As can be seen from the figure, both processes occur through an interference between two types of amplitudes, an EFT amplitude in which the Higgs boson couples to gluons, and the pure-$y_b$ amplitude in which the Higgs boson couples directly to the bottom quarks.  The two phase-space contributions which occur as part of the $\mathcal{O}(\as^2)$ corrections are separately IR finite, with the two-body term requiring UV renormalization (of $y_b$). 
 The production of the $b\overline{b}$ pair in the EFT amplitude always occurs through their coupling to the spin-1 gluon, whereas in the $y_b$ amplitude the quarks couple to the scalar Higgs. As a result, a mass term is needed to ensure a non-vanishing trace upon interference and accordingly the whole contribution will scale as 
\begin{eqnarray}
2 {\rm{Re}}(A^{\dagger}_{\rm{EFT}} A_{y_b}) \sim {y_b} \frac{m_b}{v}.
\label{eq:ybscal}
\end{eqnarray}
This scaling can be used to make the argument that these terms are effectively proportional to $y_b^2$ and are thus of the same order as the remainder of the NNLO QCD contribution. Broadly defined this argument is valid, but care must be taken at the level of the full calculation, especially in regards to the mixed renormalization scheme which is common in the literature (and we will employ here). 
While a detailed overview of our renormalization prescription is provided in section~\ref{sec:renorm}, here we simply note that, following the discussion outlined in the introduction, we work in the mixed scheme and as such  
the scaling equation (\ref{eq:ybscal}) would be written as 
\begin{eqnarray}
2 {\rm{Re}}(A^{\dagger}_{\rm{EFT}} A_{y_b}) \sim {{y^{\msbar}_b}} {{y^{\os}_b}} \, .
\end{eqnarray}
Due to the running of the bottom-quark mass ${y^{\os}_b} \sim 2{y^{\msbar}_b}(m_H)$, and as such these contributions are somewhat enhanced compared to the $(y^{\msbar}_b)^2$ which multiplies the rest of the NNLO coefficient. 

\begin{figure}
\begin{center}
\includegraphics[width=15cm]{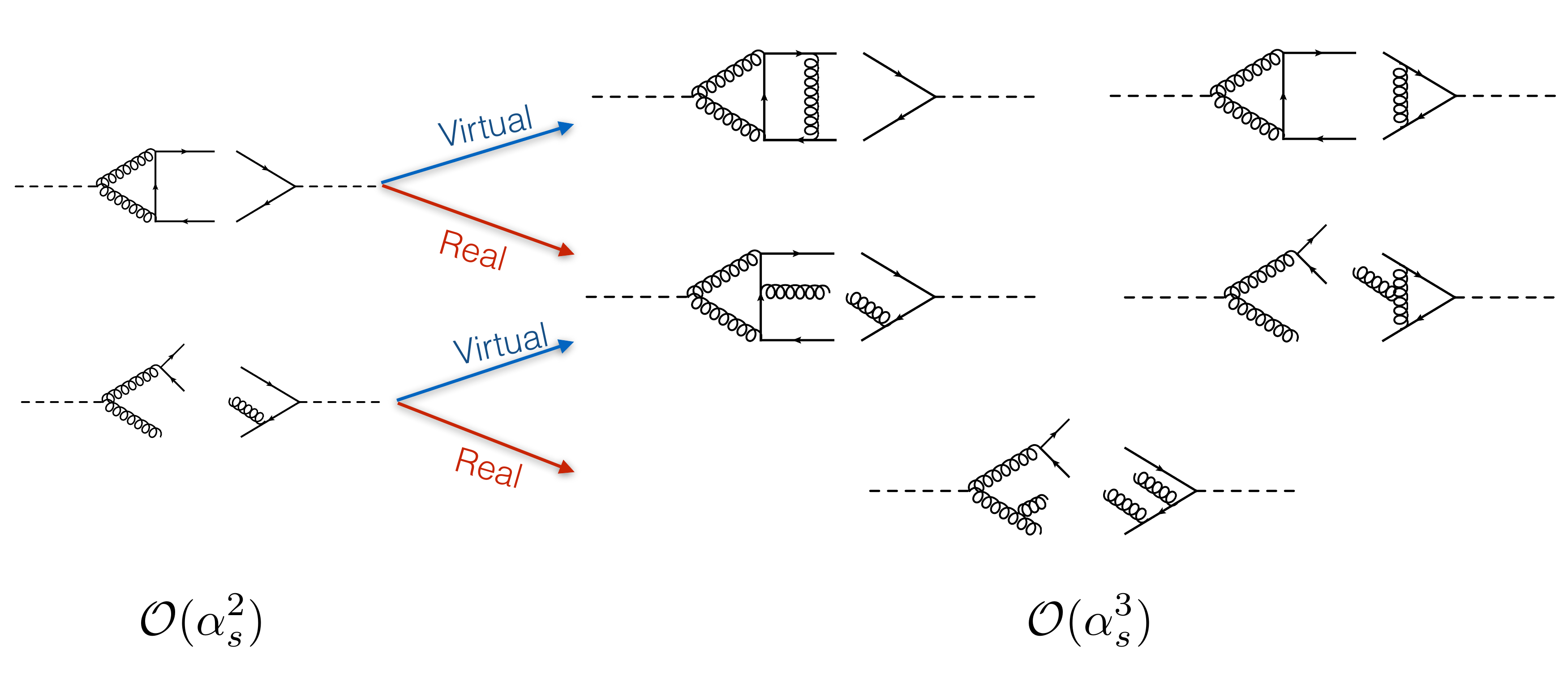}
\caption{Representative Feynman diagrams for the EFT-induced contributions at order $\mathcal{O}(\alpha_s^3)$ computed in this paper. Unlike at $\mathcal{O}(\alpha_s^2)$ (left), the different phase-space contributions are no longer separately infrared finite. The three-body phase-space contribution is particularly complicated since it has contributions which correspond to the real corrections for the two-body final-state, with the total contribution serving as the virtual corrections to the 3-body final-state. }
\label{fig:ytybas3}
\end{center}
\end{figure}

At $\mathcal{O}(\as^3)$ there are three types of phase-space contributions, for which example diagrams are presented in fig.~\ref{fig:ytybas3}. There are virtual corrections to the 2-body phase-space, and a 4-body phase-space contribution which corresponds to the real corrections to the 3-body phase-space. The 3-body phase-space itself has a rather intricate IR structure at this order. It contains both explicit IR poles arising from the one-loop integration, and implicit poles which arise when the emitted gluon becomes soft. This term therefore is more akin to a real-virtual correction in a NNLO calculation.

\subsection{Amplitude definitions and UV renormalization}
\label{sec:renorm}

The Higgs-to-partons amplitudes (required for our calculation) can be expanded in terms of the bare coefficients $C_1^B$ and $C_2^B$ (defined in eq.~\eqref{eq:LHEFT}) as follows, 
\begin{align}
\amp^B_{b \bb}\left(\as^B,m_b^{B}\right)&=y^{B}_b \, C_2^B \,  \amp_{b \bb, C_2}\left(\as^B,m_b^{B}\right)+ \frac{C^{B}_1}{v}  \amp_{b \bb, C_1}\left(\as^B,m_b^{B}\right)  \\
\amp^B_{b \bb g}\left(\as^B,m_b^{B}\right)&= y^{B}_b \, C_2^B \, \amp_{b \bb g,C_2}\left(\as^B,m_b^{B}\right)+\frac{C^{B}_1}{v} \amp_{b \bb g,C_1}\left(\as^B,m_b^{B}\right) \\
\amp^B_{b \bb ff}\left(\as^B,m_b^{B}\right)&= y^{B}_b \, C_2^B \,  \amp_{b \bb ff, C_2}\left(\as^B,m_b^{B}\right)+ \frac{C^{B}_1}{v}  \amp_{b \bb ff, C_1}\left(\as^B,m_b^{B}\right) \, ,  
\end{align}
where we recall that $C^B_1$ defines the terms which couple through the EFT vertex, and $C^B_2$ defines the bare Higgs-fermion vertex. In the four-parton decay amplitudes $ff$ can be either $gg$, $q\overline{q}$ (with $m_q=0$) or $b\bb$. The $b\bb b\bb$ amplitude exists in a separate phase-space and is both UV- and IR-finite at $\mathcal{O}(\as^3)$.

The bare parameters defined in eq.~\eqref{eq:LHEFT} require renormalization in order to make our predictions UV-finite. Following the discussion in the introduction, we work in a mixed scheme (see also for instance the discussion in ref.~\cite{Bernreuther:2004ih}) in which the couplings are defined in the $\msbar$ scheme, and the quark masses and wave functions are defined in the on-shell (OS) scheme. 
In this mixed scheme we define the renormalized quantities in terms of the bare ones as follows, 
\begin{eqnarray}
A^{B}_{\mu} &=& \sqrt{Z_A^\os} A_{\mu}\\
\as^{B} &=& Z_{\as}^{{\msbar}} \, \as \\
\psi^{B}_j &=& \sqrt{Z_{Q}^\os} \, \psi_j \\
m^{B}_j &=& Z_m^\os \, m_j \\
y_j^{B} &=& Z_{y_j}^{\msbar} \, y_j \, .
\end{eqnarray}
Expansion of the $Z^S_i$ coefficients are provided to the required accuracy in appendix~\ref{sec:appRenom}. The renormalization of the Wilson coefficients is defined as follows, 
\begin{align}
\begin{pmatrix}
C_1^{B} \\
C_2^{B} 
\end{pmatrix} = 
\begin{pmatrix}
Z_{11}^{\msbar} & 0 \\
Z_{21}^{\msbar} & 1 \\
\end{pmatrix}
\begin{pmatrix}
C_1 \\
C_2
\end{pmatrix}
\end{align}
with 
\begin{align}
Z_{11}^{\msbar}&=1+\alpha_s \frac{\partial \log(Z_{\alpha_s}^{\msbar})}{\partial \alpha_s} \\
Z_{21}^{\msbar}&=-\alpha_s \frac{\partial \log(Z_m^{\msbar})}{\partial \alpha_s} \, .
\label{eq:z21def}
\end{align}
After renormalization we can define the two-parton and three-parton decay amplitudes as follows (the four-parton amplitude does not require UV renormalization at $\mathcal{O}(\alpha_s^3)$),
\begin{align}
\begin{split}
\amp_{b \bb}\left(\as,m_b\right)&=Z_Q^\os \left[y_b \, C_2 \, Z_{y_b}^{\msbar} \,  \amp_{b \bb, C_2}\left(\as,m_b\right) \right. \\
&+ \left. \frac{C_1}{v}  \left(Z_{11}^{\msbar} \amp_{b \bb, C_1}\left(\as,m_b\right) + Z_{m}^{\msbar} \,  Z_{21}^{\msbar} \,  m_b^{\msbar} \, \amp_{b \bb, C_2}\left(\as,m_b\right) \right) \right]
\end{split}\\
\begin{split}
\amp_{b \bb g}\left(\as,m_b\right)&= Z_Q^\os\, \sqrt{Z_A^\os}\left[ y_b \, C_2 \, Z_{y_b}^{\msbar}  \, \amp_{b \bb g,C_2}\left(\as,m_b\right)\right. \\
&+ \left. \frac{C_1}{v} \left( Z_{11}^{\msbar}\amp_{b \bb g,C_1}\left(\as,m_b\right) +Z_{m}^{\msbar} \, Z^{\msbar}_{21} \,  m_b^{\msbar} \, \amp_{b \bb g,C_2} \left(\as,m_b\right) \right) \right] \, .
\end{split}
\end{align}
An unfortunate side effect of our scheme defined thus far is an inconsistency in the terms proportional to $C_1$. The two terms are separately UV divergent and are finite only when summed. The UV
poles in $\amp_{b \bb, C_1}$ and $\amp_{b \bb g, C_1}$, however,  are proportional to the on-shell mass, whereas those from $\amp_{b \bb, C_2}$ and $\amp_{b \bb g, C_2}$ have an $\msbar$ mass term arising from eq.~\eqref{eq:z21def}. While at $\mathcal{O}(\alpha_s^2)$ (which corresponds to the known literature results) this inconsistency is of limited concern, since one can switch between the pole mass and $\msbar$ mass at will, care must be taken when expanding the amplitude to $\mathcal{O}(\alpha_s^3)$. In order to consistently cancel UV poles we employ the following relationship (see for example the discussion in ref.~\cite{Melnikov:2000qh})
\begin{eqnarray}
Z_m^{\msbar} m_b^{\msbar} = Z_m^{\os}m_b \,
\end{eqnarray}
to re-write the effected terms, which  results in the following renormalized amplitude,
\begin{align}
\begin{split}
\amp_{b \bb}\left(\as,m_b\right)&=Z_Q^\os \left[y_b \, C_2 \, Z_{y_b}^{\msbar} \,  \amp_{b \bb, C_2}\left(\as,m_b\right) \right. \\
&+ \left. \frac{C_1}{v}  \left(Z_{11}^{\msbar} \amp_{b \bb, C_1}\left(\as,m_b\right) + Z_{m}^{\os} \,  Z_{21}^{\msbar} \,  m_b \, \amp_{b \bb, C_2}\left(\as,m_b\right) \right) \right]
\end{split}\\
\begin{split}
\amp_{b \bb g}\left(\as,m_b\right)&= Z_Q^\os\, \sqrt{Z_A^\os}\left[ y_b \, C_2 \, Z_{y_b}^{\msbar}  \, \amp_{b \bb g,C_2}\left(\as,m_b\right)\right. \\
&+ \left. \frac{C_1}{v} \left(Z_{11}^{\msbar} \amp_{b \bb g,C_1}\left(\as,m_b\right) +Z_{m}^{\os} \, Z^{\msbar}_{21} \,  m_b \, \amp_{b \bb g,C_2} \left(\as,m_b\right) \right) \right] \, .
\end{split}
\end{align}
Finally, we note that $C_1$ and $C_2$ must be matched to the full SM~\cite{Chetyrkin:1997un}, and therefore our calculation requires 
\begin{eqnarray}
C_1 &=& C_1^0 \left(1+\left(\frac{\as}{\pi}\right)\Delta_H^{(1)}+\mathcal{O}(\as^2)\right)
\label{eq:c1def} \\
C_2 &=& C_2^0 \left(1+\left(\frac{\as}{\pi}\right)^2\Delta_F^{(2)}+\left(\frac{\as}{\pi}\right)^3\Delta_F^{(3)}+\mathcal{O}(\as^4)\right) \, .
\label{eq:c2def}
\end{eqnarray}
Here $C_1^0 = -\as/(12\pi)$ and $C_2^0=1$, and explicit forms for the required $\Delta^{(m)}_i$ are given in appendix~\ref{sec:appRenom}.
In order to simplify formulae in the following section we introduce the following notation for the renormalized amplitudes, 
\begin{eqnarray}
\amp_{b \bb}(\alpha_s,m_b) &=&  C_2 \tilde{\amp}_{b \bb,C_2} + {C_1} \tilde{\amp}_{b \bb,C_1} \\
\amp_{b \bb g}(\alpha_s,m_b) &=&  C_2 \tilde{\amp}_{b \bb g,C_2} + {C_1}\tilde{\amp}_{b \bb g,C_1} \, ,
\end{eqnarray}
such that the renormalized $\tilde{A}$ amplitudes contain all terms apart from the overall scaling of the Wilson coefficient (recall that the $\tilde{A}_{X,C_1}$ amplitudes depend on both bare $C_1$ and $C_2$ amplitudes). When performing an expansion in $\as$ in the next section we will use the following notation to denote perturbative expansion to $\mathcal{O}(\as^j)$:
\begin{eqnarray}
\amp_{b \bb X}(\alpha_s,m_b)  &=& \sum_j \amp^{(j)}_{b \bb X} + \mathcal{O}(\alpha_s^{(j+1)}) \, ,
\end{eqnarray}
where $b \bb X$ defines the phase-space of interest, and $j$ defines the number of loops in the (unrenormalized) amplitude. We find it most convenient to refer to terms by the scaling of the unmatched coefficients $C_1^0$ and $C_2^0$ so we finally expand
\begin{eqnarray}
\amp^{(j)}_{b \bb X} &=& C_1^{0} \tilde{\amp}^{(j-1)}_{b \bb X, C_1}+ C_2^{0} \tilde{\amp}^{(j)}_{b \bb X, C_2}  \nonumber\\&& + \sum_{m=1}^{j-1} \left(\frac{\as}{\pi}\right)^m\left(   C_1^0  \Delta_H^{(m)} \tilde{\amp}^{(j-m-1)}_{b \bb X, C_1}\right) 
+\sum_{m=1}^{j}  \left(\frac{\as}{\pi}\right)^m \left(C_2^0 \Delta_F^{(m)}  \tilde{\amp}^{(j-m)}_{b \bb X, C_2}\right) \, .
\end{eqnarray}
We note that the third term in the above equation (the sum over $\Delta_H^{(m)}$) is first non-zero at $\mathcal{O}(\as^3)$.

\subsection{Partial widths for $H\rightarrow jj$}

In this section we discuss the structure of the partial widths calculated in this paper. We will study various different final-state jet requirements (0-tagged, 1-btag, 2-btag etc.), hence we refer to the process as $H\rightarrow jj$, however we stress that at no stage do we include the pure $H\rightarrow gg$ amplitude. This is because our primary interest is related to the effects of the top quark (EFT pieces) on Higgs-bottom quark physics, so we will ultimately compute rates with one or two- b-tagged jets. Therefore the reader should treat $H\rightarrow jj$ as meaning $H\rightarrow$ 2 jets, arising from a matrix element with at least one $b\bb$ pair somewhere in the diagram.  
With this caveat in mind, we define the partial width for a Higgs boson decaying to $n$ jets as follows, 
\begin{eqnarray}
\Gamma_{H\rightarrow nj} = \sum_{i=0}^{\infty} \Gamma_{H\rightarrow nj}^{(i)}.
\end{eqnarray}
At each order the partial width coefficient can be written as a sum over phase-space contributions, 
\begin{eqnarray}
\Gamma^{(i)}_{H\rightarrow nj} = \frac{1}{2m_H} \sum_{m=2}^{m_{\rm{max}}} \int |\mathcal{\tilde{A}}^{(i)}_m|^2 F_m^n(\Phi_m) d\Phi_m
\end{eqnarray}
In the above equation $m$ represents the dimension of the phase-space 
which starts at $2$ and increases to $m_{\rm{max}} = m+i$, $|\mathcal{\tilde{A}}^{(i)}_m|^2$ denotes an amplitude squared (at order $\alpha_s^i$) for the Higgs decaying to $m$ partons as discussed in the previous section.    $F_m^n(\Phi_m)$ defines the measurement function which takes $m$ partons (corresponding to a phase-space point $\Phi_m$) into $n$ jets (and applies additional requirements such as $b$-tagging) of course $F_m^n = 0$ if $m < n$. 
Here we are primarily interested in the case where the number of jets is equal to two. 
At LO and NLO only terms proportional to $C_2^0 = 1$ enter the expansion
\begin{eqnarray}
\Gamma_{H\rightarrow jj}^{(0)} = 
\frac{(C_2^0)^2}{2m_H} \int |\tilde{\amp}^{(0)}_{b \bb,C_2}|^2 F_2^2(\Phi_2) d\Phi_2
\end{eqnarray}
and
\begin{eqnarray}
\Gamma_{H\rightarrow jj}^{(1)} &=& 
\frac{(C_2^0)^2}{2m_H} \bigg( \int 2 {\rm{Re}}(\tilde{\amp}^{(0)\dagger}_{b \bb,C_2}\tilde{\amp}^{(1)}_{b \bb,C_2}) F_2^2(\Phi_2) d\Phi_2\nonumber\\&&+ 
\int |\tilde{\amp}^{(0)}_{b \bb g,C_2}|^2 F_3^2(\Phi_3) d\Phi_3 \bigg)
\end{eqnarray}
at $\mathcal{O}(\as^2)$ a term proportional to $C_1^0$ and $\Delta_F^{(1)}$ first appear, 
we define 
\begin{eqnarray}
\Gamma_{H\rightarrow jj}^{(2)} =\delta\Gamma_{H\rightarrow jj}^{C_2^2,(2)}+\delta\Gamma_{H\rightarrow jj}^{C_2C_1,(0)}
+\delta\Gamma_{H\rightarrow jj}^{\Delta^{{(2)}}_F\,(0)}
\end{eqnarray}
where 
\begin{eqnarray}
\delta\Gamma_{H\rightarrow jj}^{C_2^2,(2)}&=& 
\frac{(C_2^0)^2}{2m_H} \bigg( \int \left(2 {\rm{Re}}(\tilde{\amp}^{(0)\dagger}_{b \bb,C_2}\tilde{\amp}^{(2)}_{b \bb,C_2}) + |\tilde{A}^{(1)}_{b\bb,C_2}|^2\right)F_2^2(\Phi_2) d\Phi_2\nonumber\\&&
+ \int 2{\rm{Re}}(\tilde{\amp}^{(0)\dagger}_{b \bb g,C_2}\tilde{\amp}^{(1)}_{b \bb g,C_2}) F_3^2(\Phi_3) d\Phi_3 
+ \sum_{ff = gg,\,q\overline{q}}\int |\tilde{\amp}^{(0)}_{b \bb ff,C_2}|^2 F_4^2(\Phi_4) d\Phi_4 \nonumber\\&& + \int |\tilde{\amp}^{(0)}_{b \bb b\bb,C_2}|^2 F_4^2(\Phi^m_4) d\Phi^m_4   \bigg).
\end{eqnarray}
These pieces correspond to the double-virtual, real-virtual and real-real pieces of the NNLO coefficient. We note that we have separated the $H\rightarrow 4b$ phase-space $(\Phi^m_4)$ from the remainder of the four-body phase-space.  For a recent discussion of their calculation including mass-effects we refer the reader to refs.~\cite{Bernreuther:2018ynm,Behring:2019oci}. The pieces sensitive to the top quark are defined as 
\begin{eqnarray}
\delta\Gamma_{H\rightarrow jj}^{C_2C_1,(0)} =  
\frac{(C^0_2 C^0_1)}{2m_H} \bigg(&&\int 2 {\rm{Re}}(\tilde{\amp}^{(0)\dagger}_{b \bb,C_2}\tilde{\amp}^{(1)}_{b \bb,C_1})F_2^2(\Phi_2) d\Phi_2\nonumber \\ 
+
&&\int 2 {\rm{Re}}(\tilde{\amp}^{(0)\dagger}_{b \bb g,C_2}\tilde{\amp}^{(0)}_{b \bb g,C_1})F_3^2(\Phi_3) d\Phi_3\bigg)
\label{eq:as2c2c1}
\end{eqnarray}
In this paper eq.~(\ref{eq:as2c2c1}) defines our ``LO''. In reality there are effectively two LO contributions, since the two-body and three-body final states have different IR factorization properties. The final contribution at $\as^2$ is simply a rescaling of the $\as^0$ term arising from matching the EFT to the SM: 
\begin{eqnarray}
\delta\Gamma_{H\rightarrow jj}^{\Delta_F^{{(2)}},(0)}=
\left(\frac{\as}{\pi}\right)^2\Delta_F^{{\rm{(2)}}}\Gamma_{H\rightarrow jj}^{(0)}
\end{eqnarray}
We now turn our attention to the primary focus of this paper, which is the $\as^3$ coefficient of the $H\rightarrow b\bb$ partial width, 
\begin{eqnarray}
\Gamma_{H\rightarrow jj}^{(3)} =\delta\Gamma_{H\rightarrow jj}^{C_2^2,(3)}+\delta\Gamma_{H\rightarrow jj}^{C_2C_1,(1)}
+\delta\Gamma_{H\rightarrow jj}^{C_1^2}
+\delta\Gamma_{H\rightarrow jj}^{\Delta^{{\rm{(2)}}}_F,(1)}
+\delta\Gamma_{H\rightarrow jj}^{\Delta^{{\rm{(3)}}}_F}
+\delta\Gamma_{H\rightarrow jj}^{\Delta^{{\rm{(1)}}}_H}
\label{eq:as3bdown}
\end{eqnarray}
The coefficient $\delta\Gamma_{H\rightarrow jj}^{C_2^2,(3)}$ has by far the most intricate 
IR structure, with an expansion up to the triple-real phase-space with $H\rightarrow 5$ partons. It was studied in detail in ref.~\cite{Mondini:2019gid} (in the massless approximation) and for brevity we do not provide an expansion of this term here. The remaining terms in eq.~\eqref{eq:as3bdown} have not yet been fully computed (and implemented into a fully flexible Monte Carlo code), however we note that the most technically complex part of the remaining pieces, corresponding to the two-loop EFT amplitudes were recently computed in ref.~\cite{Anastasiou:2020qzk}. 
The results presented in this paper, combined with those presented in ref.~\cite{Mondini:2019gid} allow for a complete differential prediction for $H\rightarrow jj$ at  $\mathcal{O}(\as^3)$ up to (suppressed) mass-effects in the $\delta\Gamma_{H\rightarrow jj}^{C_2^2,(3)}$ term (and presumed small top-quark mass effects missed by the EFT). 
Let us investigate the remaining terms in eq.~\eqref{eq:as3bdown} in more detail, firstly  we have the ``NLO'' correction to the mixed $C_2 C_1$ term which appeared in the $\alpha_s^2$ coefficient, 
\begin{eqnarray}
\delta\Gamma_{H\rightarrow jj}^{C_2C_1,(1)} =
\frac{(C^0_2 C^0_1)}{2m_H} \bigg(&&\int 2 {\rm{Re}}(\tilde{\amp}^{(0)\dagger}_{b \bb,C_2}\tilde{\amp}^{(2)}_{b \bb,C_1}+\tilde{\amp}^{(1)\dagger}_{b \bb,C_2}\tilde{\amp}^{(1)}_{b \bb,C_1})F_2^2(\Phi_2) d\Phi_2\nonumber \\ 
+
&&\int 2 {\rm{Re}}(\tilde{\amp}^{(0)\dagger}_{b \bb g,C_2}\tilde{\amp}^{(1)}_{b \bb g,C_1}+\tilde{\amp}^{(1)\dagger}_{b \bb g,C_2}\tilde{\amp}^{(0)}_{b \bb g,C_1})F_3^2(\Phi_3) d\Phi_3 \nonumber\\
+&&\sum_{ff = gg,\,q\overline{q}}\int 2{\rm{Re}}(\tilde{\amp}^{(0)\dagger}_{b \bb ff,C_2}\tilde{\amp}^{(0)}_{b \bb ff,C_1}) F_4^2(\Phi_4) d\Phi_4 \nonumber\\
 +&&\int 2{\rm{Re}}(\tilde{\amp}^{(0)\dagger}_{b \bb b\bb,C_2}\tilde{\amp}^{(0)}_{b \bb b\bb,C_1}) F_4^2(\Phi^{m}_4) d\Phi^{m}_4\bigg) .
\label{eq:as3c2c1}
\end{eqnarray}
At $\as^3$, a term which is not proportional to $C_2^0$ appears for the first time, 
\begin{eqnarray}
\delta\Gamma_{H\rightarrow jj}^{C^2_1} =
\frac{(C^0_1)^2}{2m_H} \bigg(&&\int |\tilde{\amp}^{(0)}_{b \bb g,C_1}|^2F_3^2(\Phi_3) d\Phi_3 \bigg).
\end{eqnarray}
Finally, terms arising from matching the EFT to the SM are defined as follows, 
\begin{eqnarray}
\delta\Gamma_{H\rightarrow jj}^{\Delta_F^{{\rm{(2)}}},(1)}&=&
\left(\frac{\as}{\pi}\right)^2\Delta_F^{{\rm{(2)}}}\Gamma_{H\rightarrow jj}^{(1)} \\
\delta\Gamma_{H\rightarrow jj}^{\Delta_F^{{\rm{(3)}}}}&=&
\left(\frac{\as}{\pi}\right)^3\Delta_F^{{\rm{(3)}}}\Gamma_{H\rightarrow jj}^{(0)} \\
\delta\Gamma_{H\rightarrow jj}^{\Delta_H^{{\rm{(1)}}}}&=&
\left(\frac{\as}{\pi}\right)\Delta_H^{{\rm{(1)}}}\delta\Gamma_{H\rightarrow jj}^{C_2C_1,(0)}.
\end{eqnarray}
We note that each term here is individually $\mathcal{O}(\as^3)$and the prefactor is related to the expansion in terms of $\Delta_F$ and $\Delta_H$.

\subsection{IR divergences}

For a suitably-inclusive definition of the measurement function appearing in the previous equations, the (total) partial widths defined are IR finite. However, care must be taken if more differential quantities are considered, since the individual phase-space terms can exhibit singularities which only cancel upon integration over the unresolved regions of phase-space. In eq.~(\ref{eq:as3bdown}) the following terms contain individual pieces which are IR divergent, 
$\delta\Gamma_{H\rightarrow jj}^{C_2^2,(3)}$, $\delta\Gamma_{H\rightarrow jj}^{C_2C_1,(1)}$ and  $\delta\Gamma_{H\rightarrow jj}^{\Delta^{{\rm{(2)}}}_F,(1)}$. Whereas, 
$\delta\Gamma_{H\rightarrow jj}^{C_1^2}$,
$\delta\Gamma_{H\rightarrow jj}^{\Delta^{{\rm{(3)}}}_F}$, and $\delta\Gamma_{H\rightarrow jj}^{\Delta^{{\rm{(1)}}}_H}$ are IR finite in all phase-space configurations. As mentioned previously, of the IR sensitive terms, $\delta\Gamma_{H\rightarrow jj}^{C_2^2,(3)}$ is by far the most complicated since it exhibits triple unresolved limits. In ref.~\cite{Mondini:2019gid} (for massless $b$ quarks) these limits where regulated using a combination of the projection-to-Born technique~\cite{Cacciari:2015jma}, $N$-jettiness slicing~\cite{Gaunt:2015pea,Boughezal:2015dva}, and Catani-Seymour dipole subtraction~\cite{Catani:1996vz}. Combined with the results for the fully inclusive width at this order~\cite{Chetyrkin:1996sr} allowed for the construction of a fully-differential Monte Carlo code. In this paper only $\delta\Gamma_{H\rightarrow jj}^{C_2C_1,(1)}$ and  $\delta\Gamma_{H\rightarrow jj}^{\Delta^{{\rm{LO}}}_F,(1)}$ require IR regulation. Both of these terms have the same singular structure as a traditional NLO calculation, although $\delta\Gamma_{H\rightarrow jj}^{C_2C_1,(1)}$ is complicated by the presence of two different ``LO'' terms, and as a result different pieces have different factorization properties.  Since the structure is similar to a NLO calculation, we use Catani-Seymour dipoles, extended to include massive partons~\cite{Catani:2002hc,Dittmaier:1999mb}. In order to make our predictions fully-differential in all IR-safe observables we modify $\delta\Gamma_{H\rightarrow jj}^{C_2C_1,(1)}$ as follows, 
\begin{eqnarray}
\delta\Gamma_{H\rightarrow jj}^{C_2C_1,(1)} =
\frac{(C^0_2 C^0_1)}{2m_H} \bigg(&&\int 
\bigg\{ 2 {\rm{Re}}\left(\tilde{\amp}^{(0)\dagger}_{b \bb,C_2}\tilde{\amp}^{(2)}_{b \bb,C_1}+\tilde{\amp}^{(1)\dagger}_{b \bb,C_2}\tilde{\amp}^{(1)}_{b \bb,C_1}\right)
\nonumber\\+&&2 \sum_{ij}\tilde{D}_{ij}  \times {\rm{Re}}\left(\tilde{\amp}^{(0)\dagger}_{b \bb,C_2}\tilde{\amp}^{(1)}_{b \bb,C_1}\right) \bigg\}F_2^2(\Phi_2) d\Phi_2\nonumber \\ 
+
&&\int \bigg\{ 2 {\rm{Re}}(\tilde{\amp}^{(0)\dagger}_{b \bb g,C_2}\tilde{\amp}^{(1)}_{b \bb g,C_1}+\tilde{\amp}^{(1)\dagger}_{b \bb g,C_2}\tilde{\amp}^{(0)}_{b \bb g,C_1})F_3^2(\Phi_3)  \nonumber\\
+&& 2 \sum_{ij}\tilde{D}_{ij}  \times {\rm{Re}}\left(\tilde{\amp}^{(0)\dagger}_{b \bb g,C_2}\tilde{\amp}^{(0)}_{b \bb g,C_1}\right)F_3^2(\Phi_3)
\nonumber\\
- &&2 \sum_{i,j,k} D_{i,j,k}(\Phi_{ijk}) \times {\rm{Re}}\left(\tilde{\amp}^{(0)\dagger}_{b \bb,C_2}\tilde{\amp}^{(1)}_{b \bb,C_1}\right)F_2^2(\hat{\Phi}_2)  \bigg\}d\Phi_3 \nonumber\\
+&&\bigg\{\int 2\sum_{ff = gg,\,q\overline{q}} {\rm{Re}}(\tilde{\amp}^{(0)\dagger}_{b \bb ff,C_2}\tilde{\amp}^{(0)}_{b \bb ff,C_1}) F_4^2(\Phi_4)  \nonumber\\
-&&2 \sum_{i,j,k} \bigg[ D_{i,j,k}(\Phi_{ijk}) \times {\rm{Re}}\left(\tilde{\amp}^{(0)\dagger}_{b \bb g,C_2}\tilde{\amp}^{(0)}_{b \bb g,C_1}\right) \nonumber \\+&& D^z_{i,j,k}(\Phi_{ijk}) \times {\rm{Re}}\left(\tilde{\amp}^{(0)\dagger}_{b \bb n,C_2}\tilde{\amp}^{(0)}_{b \bb n,C_1}\right) \bigg]
F_3^2(\hat{\Phi}_3)
\bigg\}d\Phi_4 \nonumber\\
 +&&\int 2{\rm{Re}}(\tilde{\amp}^{(0)\dagger}_{b \bb b\bb,C_2}\tilde{\amp}^{(0)}_{b \bb b\bb,C_1}) F_4^2(\Phi^{m}_4) d\Phi^{m}_4\bigg) \, . 
\label{eq:as3c2c1IR}
\end{eqnarray}
In eq.~(\ref{eq:as3c2c1IR}) the subtraction terms are defined by $D_{i,j,k}(\Phi_{ijk})$ where $\Phi_{ijk} = \{p_i,p_j,p_k\}$ defines the emitter, emitted and spectator partons and $\hat{\Phi}$ defines the mapped phase-space used in the lower order subtraction (and relevant measurement function). Integrating the dipoles over the dipole phase-space generates the integrated dipole terms $\tilde{D}_{ij}$. Due to the low total number of total partons, color conservation simplifies the color correlations ${\bf{T}}_i\cdot {\bf{T}}_j$ to the level of overall multiplicative factors. However spin correlations in $g\rightarrow gg$ and $g\rightarrow q\overline{q}$ splitting require care, and are regulated additionally by the azimuthal terms $D^{z}_{i,j,k}$
(see ref.~\cite{Catani:2002hc} for further details), which require modified tree-level amplitudes $\tilde{\amp}^{(0)}_{b \bb n,C_i}$ with the gluon polarization vector replaced by the spin-reference vector $n$. 
In eq.~(\ref{eq:as3c2c1IR}) the IR structure of the two-body and four-body phase-space is self-evident, the two-body phase-space only has explicit poles in $\epsilon$, which are directly canceled by the integrated terms arising from the tree-body subtraction term. The four-body phase-space only has implicit poles which arise upon integration over the unresolved regions of phase-space. These are compensated by the subtraction terms on the seventh and eighth lines of the formula. The three-body phase-space is more complicated since it requires regulation of both types of term. The fourth line corresponds to the integrated subtractions of the four-body phase-space, these cancel the explicit poles in $\epsilon$ which reside in both $\tilde{\amp}^{(0)\dagger}_{b \bb g,C_2}\tilde{\amp}^{(1)}_{b \bb g,C_1}$ and $\tilde{\amp}^{(1)\dagger}_{b \bb g,C_2}\tilde{\amp}^{(0)}_{b \bb g,C_1}$. The fifth line in eq.~(\ref{eq:as3c2c1IR}) defines the counter-terms required to cancel the implicit poles in the ($\epsilon^0$) part of $\tilde{\amp}^{(0)\dagger}_{b \bb g,C_2}\tilde{\amp}^{(1)}_{b \bb g,C_1}$ (which we recall corresponds the the one-loop EFT amplitude interfered with the tree-level $H\rightarrow b\bb g$). The 
other term $\tilde{\amp}^{(1)\dagger}_{b \bb g,C_2}\tilde{\amp}^{(0)}_{b \bb g,C_1}$ which corresponds to a tree-level EFT amplitude interfered with a one-loop $H\rightarrow b\bb g$ amplitude (in which the Higgs couples directly to the $b$ quarks) is finite in the limit $p_g \rightarrow 0$. This can be seen since although the one-loop amplitude exhibits soft singular behaviour, it is sufficiently damped by the EFT tree (which requires a hard emission) to avoid any further singular behavior. 

The IR behaviour of $\delta\Gamma_{H\rightarrow jj}^{\Delta^{{\rm{(2)}}}_F,(1)}$ is exactly that of the NLO computation of the partial width and can be readily expressed in terms of the dipole functions listed above. For brevity we do not provide the result here. 

\subsection{Calculation methodologies}

All amplitudes calculated in this publication were generated with QGraf~\cite{Nogueira:1991ex} and then interfered with each other to obtain the relevant squared amplitudes. Traces were computed with TRACER~\cite{Jamin:1991dp} and the resulting integrals reduced to master integrals with the help of integration-by-parts identities generated by {\tt LiteRed}~\cite{Lee:2012cn,Lee:2013mka} and {\tt Reduze}~\cite{vonManteuffel:2012np}. The expressions for the relevant two-loop master integrals have been presented in~\cite{Mastrolia:2017pfy,DiVita:2019lpl}, where required, one-loop integrals were evaluated numerically using {\tt QCDLoop}~\cite{Ellis:2007qk,Carrazza:2016gav}.

\subsection{Additional master integrals for \hcc}

\begin{figure}
\begin{center}
\includegraphics[width=12cm]{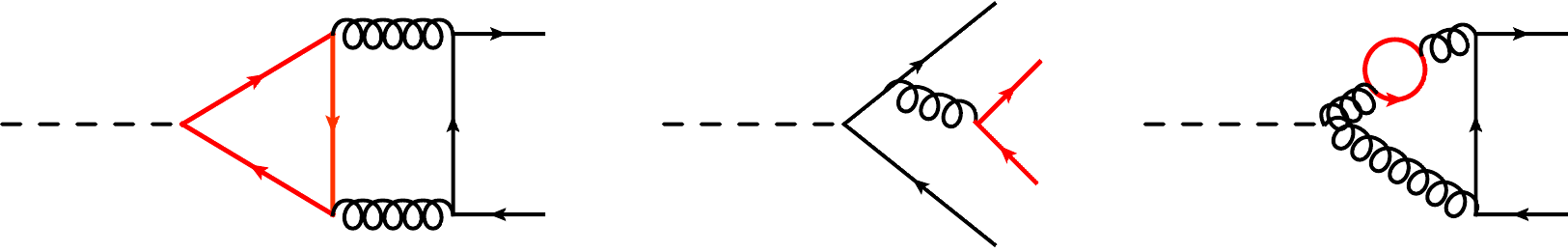}
\caption{Additional contributions which can arise in \hcc with the addition of a second active heavy quark (bottom) shown in red.}
 \label{fig:charmfigs}
\end{center}
\end{figure}

Thus far, the discussion in this section has specifically related to the computation of \hbb. The calculation can, however, be readily extended to include the decay \hcc.
The primary difference when considering charm quarks is that there are now two active massive flavors (charm and bottom quarks).
The vast majority of the calculation can simply be obtained from the results in this section with the replacements $b\rightarrow c$, $m_b \rightarrow m_c$ and $y_b \rightarrow y_c$. 
 We also define the replacement $C_2 \rightarrow C_3$ (this is purely for later notational connivence $C_3=C_2$). 
 This leaves contributions which are dependent on the bottom and charm quark simultaneously. Example topologies are shown in fig.~\ref{fig:charmfigs}. We see there are three basic types of additional corrections. Firstly there are two body contributions which are proportional to $y_b y_c$, while a second type of term contains the four-quark amplitude $H\rightarrow b\overline{b}c\overline{c}$. Depending on which quark line the Higgs couples to, the resultant matrix element squared can be proportional to either $y_c^2$, $y_b y_c$ or $y_b^2$. 
 There is therefore some ambiguity as to whether these pieces are included in \hbb or \hcc (or both). 
 In our computation of \hbb we take $m_c=0$ and these terms enter as part of the double-real NNLO correction (keeping the $y_b^2$ term only). 
In our computation of \hcc we include all three types of contribution as a component of the partial width. 
 These $y_b$ sensitive pieces, which occur at $\mathcal{O}(\alpha_s^2)$ are therefore included in our calculation. However, their impact will be shown to be small, as such we do not compute the $\mathcal{O}(\alpha_s^3)$ correction to these processes in this paper. 
A final type of diagram, which also occurs at $\mathcal{O}(\alpha_s^3)$, is also shown in fig.~\ref{fig:charmfigs}. This term induces an $m_b$ dependence into the EFT-$y_c$ piece. 

The new loop amplitudes discussed above require 20 new additional master integrals (MI's). These integrals have been computed for the case of bottom-top loops in~\cite{Primo:2018zby}, however the results are not fully public. In this work we have reproduced the calculation, using the method of differential equations~\cite{Kotikov:1990kg,Remiddi:1997ny,Gehrmann:1999as} and have included the results as ancillary material in the {\tt{arXiv}} submission. We outline the details of the computation in the remainder of this section, while enthusiastic readers can find further technical details in appendix~\ref{app:MI}. 
We define the kinematics of the process as follows, 
 $H(p_H)\rightarrow c(p_1)+\bar{c}(p_2)$ where $p_1^2=p_2^2=m_c^2$ and $p_H^2=m_H^2$. The additional master integrals can be expressed in terms of the following integral family
\begin{align}
\mathcal{I}(n_1,\dots,n_7)=\left(\frac{1}{\Gamma(1+\eps)} \right)^2 \left(\frac{\mu^2}{m_b^2} \right)^{2\eps} \int \frac{d^D k_1} {\left(i \pi \right)^{D/2}} \frac{d^D k_2} {\left(i \pi \right)^{D/2}} \frac{1}{D_1^{n_1} \dots D_7^{n_7}} \, ,
\end{align}
with the dimensional regularization parameter $\eps=\frac{D-4}{2}$ and the propagators defined as
\begin{gather}
D_1 = k_1^2-m_c^2,\quad
D_2 = k_2^2-m_b^2,\quad
D_3 = (k_1+p_1)^2,\quad
D_4 = (k_1-k_2+p_1)^2-m_b^2, \nn
D_5 = (k_1-p_2)^2,\quad
D_6 = (k_2-p_H)^2-m_b^2,\quad
D_7 = (k_2+p_1)^2 \,. 
\end{gather}
It is further convenient to introduce the following dimensionless variables 
\begin{align}
- \frac{m_H^2}{m_b^2}=\frac{(1-w^2)^2}{w^2}\, , \ \ \frac{m_c^2}{m_b^2}=\frac{(1-w^2)^2 \, z^2}{(1-z^2)^2 \, w^2} \, .
\end{align}
In order to determine the MI's we use the Magnus algorithm~\cite{Argeri:2014qva,DiVita:2014pza} which allows us to identify a particular set of integrals defined in~\eqref{def:CanBasis}, which satisfy a canonical differential equation
\begin{equation}
d \GGvec = \eps \sum_{i=1}^{12} \MM_i  \dlog(\eta_i) \GGvec \, .
\label{eq:canonicalDEQ}
\end{equation}
Here $\MM_i$ is a $20\times 20$ matrix of rational numbers. For a canonical differential equation the dimensional regularization parameter $\eps$ has factorized from the kinematics, which are encoded in a $\dlog$-form. The arguments of these $\dlog$'s are called letters, and taken together they form the alphabet of our problem
 \begin{align}
 \begin{alignedat}{4}
\eta_1 & =1-w\,,&\quad
\eta_2 & =w\,, & \quad 
\eta_3 & =1+w\,, & \quad
\eta_4 & =1+w^2\,,  \nn
\eta_5 & =1-z\,,&\quad
\eta_6 & =z\,,&\quad
\eta_7 & =1+z\, ,& \quad
\eta_8 & =1+z^2 \, , \nn
\eta_9 & =w-z  \, ,& \quad
\eta_{10} & =w+z \, ,& \quad
\eta_{11} & =1-w\,z  \, ,& \quad
\eta_{12} & =1+w\,z  \, . 
\end{alignedat} \stepcounter{equation}\tag{\theequation}
\label{alphabet}
 \end{align} 

Since all letters in our alphabet exhibit algebraic roots, the solution of the differential equation can be expressed in terms of generalized polylogartihms~\cite{Goncharov:polylog,Remiddi:1999ew,Gehrmann:2001jv,Vollinga:2004sn}. To complete the solutions of the differential equations a boundary constant has to be specified. In our case demanding the regularity of our solutions at the pseudothreshold $m_H^2=0$ provides relations between the boundary constants of 17 integrals. The remaining three integrals $\GG_{1,2,5}$ are then taken as an independent input. Their solutions were first computed in the Euclidean region $m_H^2<0, m_c^2>0, m_b^2>0$ and then analytically continued to the production region $m_H^2>4m_c^2$ by adding a small positive imaginary part to the Higgs mass $m_H^2 \rightarrow m_H^2 +i 0^+$. We have checked our solutions in both regions against their numerical expression provided by {\tt SecDec}~\cite{Borowka:2012yc,Borowka:2015mxa} and found full agreement. 

\section{Results for $H\rightarrow b\bb$}

\label{sec:results}
\begin{table}
\begin{center}
\def\arraystretch{1.7}
\begin{tabular}{|c |c || c c c |} 
\hline
& Partial Width [MeV] & $\mu = m_H/2$ & $\mu = m_H$ & $\mu = 2 m_H$ \\
\hline
$\mathcal{O}(\as^0)$ & $\Gamma^{(0)}_{H\rightarrow jj}$ &2.1599 & 1.9180 & 1.7246  \\
$\mathcal{O}(\as^1)$ & $\Gamma^{(1)}_{H\rightarrow jj}$ &0.2603 &  0.3989 &  0.4822\\
\hline 
\hline
$\mathcal{O}(\as^2)$ &  $\delta\Gamma_{H\rightarrow jj}^{C_2^2}$   & -0.0099  & 0.0732 &   0.1418  \\
& $\delta\Gamma_{H\rightarrow jj}^{C_1C_2}$  &0.02656  & 0.02418 & 0.02202   \\
\hline 
\hline
$\mathcal{O}(\as^3)$ &  $\delta\Gamma_{H\rightarrow jj}^{C_2^2}$   & -0.01514  & 0.00431   &  0.03542  \\
&$\delta\Gamma_{H\rightarrow jj}^{C_1C_2}$  & 0.00946 & 0.01306 & 0.01476   \\
&$\delta\Gamma_{H\rightarrow jj}^{C_1^2}$  & 0.00920 & 0.00670 &   0.00504 \\
\hline 
\end{tabular}
\caption{Inclusive partial width results for \hbb at LO, and higher order coefficients for NLO $y_b^2$ contributions, and the top induced pieces at $\mathcal{O}(\alpha_s^2)$ and $\mathcal{O}(\alpha_s^3)$.} 
\label{tab:incybyt}
\end{center}
\end{table}
We have implemented the calculation described in the previous section into a fully flexible parton-level Monte Carlo code, based upon the structure of MCFM~\cite{Campbell:1999ah,Campbell:2011bn,Campbell:2015qma}. 
For the results presented in this paper we take the mass of the decaying Higgs boson to be $m_H = 125$ GeV, we take $\alpha_s(m_Z) = 0.118$. 
We use a pole mass of $m_b=4.78$ GeV, and in the Yukawa coupling the $\msbar$ mass
of $m_b(m_H)=2.80$ GeV.  Our remaining electroweak parameters are $G_F =0.116638 \times 10^{-4}$ GeV$^{-2}$, and $m_W =80.385$ GeV, and $v=246$ GeV. 

Since we are interested in presenting results with the application of jet clustering and b-tagging, we cluster partons into jets using the Durham jet algorithm~\cite{Brown:1990nm,Catani:1991hj}. 
This algorithm starts from a set of ordered momentum (which for us correspond to partons), and computes the following quantity $y_{ij}$ for all pairs of objects $i$ and $j$:
\begin{eqnarray}
y_{ij} = \frac{2\,{\rm{min}}(E^2_i,E^2_j)(1-\cos{\theta_{ij}})}{Q^2}.
\end{eqnarray} 
Where $E_i$ is the energy of object  $i$, and $\theta_{ij}$ defines the angle between the two objects $i$ and $j$.
$Q$ is the total energy of the system, which in our case is $m_H$. If $y_{ij} < y_{\rm{cut}}$, the two objects are combined into a new one with four momentum equal to $p_i+p_j$. The procedure is then iterated until no more clustering is possible and the resulting $n$ final objects are classified as $n$ jets. We work in the Higgs rest frame, although in order to define dynamic observables at LO we define the fictitious collision axis $\pm{\hat{z}}$ which allows us to compute rapidities and transverse momentum of jets at LO.  

We begin by computing inclusive partial widths, 
our results are summarized in Table~\ref{tab:incybyt}. We list the coefficients of the partial width expansion up to third order. 
Since the primary interest of this work is the relative size of the pieces proportional to $C_1$ we present the following breakdown at each order (where appropriate)
\begin{eqnarray}
\delta \Gamma_{H\rightarrow jj}^{C_2^2} &=& \bigg\{
\begin{array}{c}
\delta \Gamma_{H\rightarrow jj}^{C_2^2,(2)} + \delta \Gamma_{H\rightarrow jj}^{\Delta_F^{(2)},(0)}  \quad {\rm{at}} \quad  \alpha_s^2 \\
 \delta \Gamma_{H\rightarrow jj}^{C_2^2,(3)} + \delta \Gamma_{H\rightarrow jj}^{\Delta_F^{\rm{(2)}},(1)}+ \delta \Gamma_{H\rightarrow jj}^{\Delta_F^{\rm{(3)}},(0)}  \quad {\rm{at}} \quad  \alpha_s^3\, .
 \end{array} 
 \end{eqnarray}
 These pieces define those which scale like $y_b^2$, on the other hand, the top sensitive pieces are grouped at each order as follows, 
 \begin{eqnarray}
 \delta \Gamma_{H\rightarrow jj}^{C_1C_2} &=& \bigg\{
 \begin{array}{c}
 \delta \Gamma_{H\rightarrow jj}^{C_1C_2,(0)} \quad {\rm{at}} \quad  \alpha_s^2 \\
\delta \Gamma_{H\rightarrow jj}^{C_1C_2,(1)} + \delta \Gamma_{H\rightarrow jj}^{\Delta_H^{\rm{(1)}},(1)}  \quad {\rm{at}} \quad  \alpha_s^3 \, .\\
 \end{array}
 \end{eqnarray}
 In the discussion in the rest of this section we will refer to these combinations as EFT-$y_b$ pieces. 
 At $\as^3$ we also separate the contribution arising from $C_1^2$ (referred to subsequently as EFT$^2$ pieces).
 In the computation of the $\delta \Gamma_{H\rightarrow jj}^{C_2^2,(3)}$ and $\delta \Gamma_{H\rightarrow jj}^{C_2^2,(2)}$ contributions we take the inclusive partial width~\cite{Chetyrkin:1996sr}  with $m_b=0$ and re-weight by the LO with the full $m_b$-kinematics. 

Inspection of the table confirms that the mixed EFT-$y_b$ pieces contribute around a percent to the partial width at $\mathcal{O}(\alpha_s^2)$. The main results of this paper are included in the $\mathcal{O}(\alpha_s^3)$ rows. We observe that the mixed EFT-$y_b$ has a sizable correction at NLO (of which around 1/3 is made up of the correction arising from the Wilson coefficient), the ``LO'' to ``NLO'' $K$-factor for this process is around 1.3-1.5 for the scales presented in the table. The EFT$^2$ contribution is of the same order as the mixed term (and $y_b^2$ coefficient at this order), and contributes around 0.5\% to the total partial width. Rescaling $y_b = \kappa_b y_b$ and $C^0_1 = \kappa_g C^0_1$ we can write 
the total partial width (at $\mu=m_H$) as follows, 
\begin{eqnarray}
\Gamma^{{\rm{N3LO}}}_{H\rightarrow b\overline{b} +X}[\mu=m_H] =  2.394 \times \left(\kappa_b^2 + 0.016 \times \kappa_b \kappa_g + 0.003 \times \kappa_g^2 \right) + \mathcal{O}(\alpha_s^4) \quad {\rm{MeV}}
\end{eqnarray}
Thus the effects of having $\kappa_g\ne 0$ in the above equation change the extraction of the partial width by around 2\%. Given the intricate phase-space of these terms we expect that there is a strong sensitivity to jet-clustering and selection criteria which may be employed in an experimental analysis. We investigate this in the next section. 


\subsection{Differential predictions for {\hbb}}

In the last section we presented results for the EFT contributions to the inclusive width at $\mathcal{O}(\alpha_s^3)$, of course, the primary advantage of our calculation is the ability to employ arbitrary jet clustering and cuts to the decay products. Indeed, the impact of the EFT initiated pieces are expected to be very sensitive to the precise nature of the final state selection cuts. This can already been seen at the $\mathcal{O}(\alpha_s^2)$ level, approximately 80\% of the mixed $C_1 C_2$ coefficient arises from the 3-body decay channel. Therefore any selection requirement which eliminates 3-jet topologies (for instance running the Durham-jet algorithm with a small $y_{\rm{cut}}$, so that more events are classified as three-jet events, then asking for $H\rightarrow jj$) will dramatically curtail the contribution from the mixed pieces. A further sensitivity to the final state definition arises through the application of $b$-tagging. Since the EFT contributions produce the final state bottom quarks through $g\rightarrow b\overline{b}$ splitting, many favorable phase-space configurations will result in jets with both $b$ and $\bb$ partons placed in the same jet. Therefore requiring two resolved $b$-jets will damp the interference terms. 
With these points in mind we therefore compare predictions with the following three jet clustering requirements:
\begin{itemize}
    \item 
    No veto on additional jets
    \item 
    Requiring exactly two jets
    \item 
    Requiring exactly two $b$-jets. 
\end{itemize}
We take $y_{\rm{cut}} = 0.1$ as a representative jet clustering choice (assuming these jets are somewhat similar to an anti-$k_T$ jet with $p_T \sim y_{cut} m_H$). Since we retain the mass of the bottom quark throughout we are able to classify jets containing a $b\overline{b}$ pair as $b-$tagged. However, as mentioned before, requiring two $b$-tags effectively vetoes these configurations such that differences arising from the treatment of $g\rightarrow b\overline{b}$ (which can be troublesome for massless calculations) are of limited concern here. 

\begin{figure}
\begin{center}
\includegraphics[width=8cm]{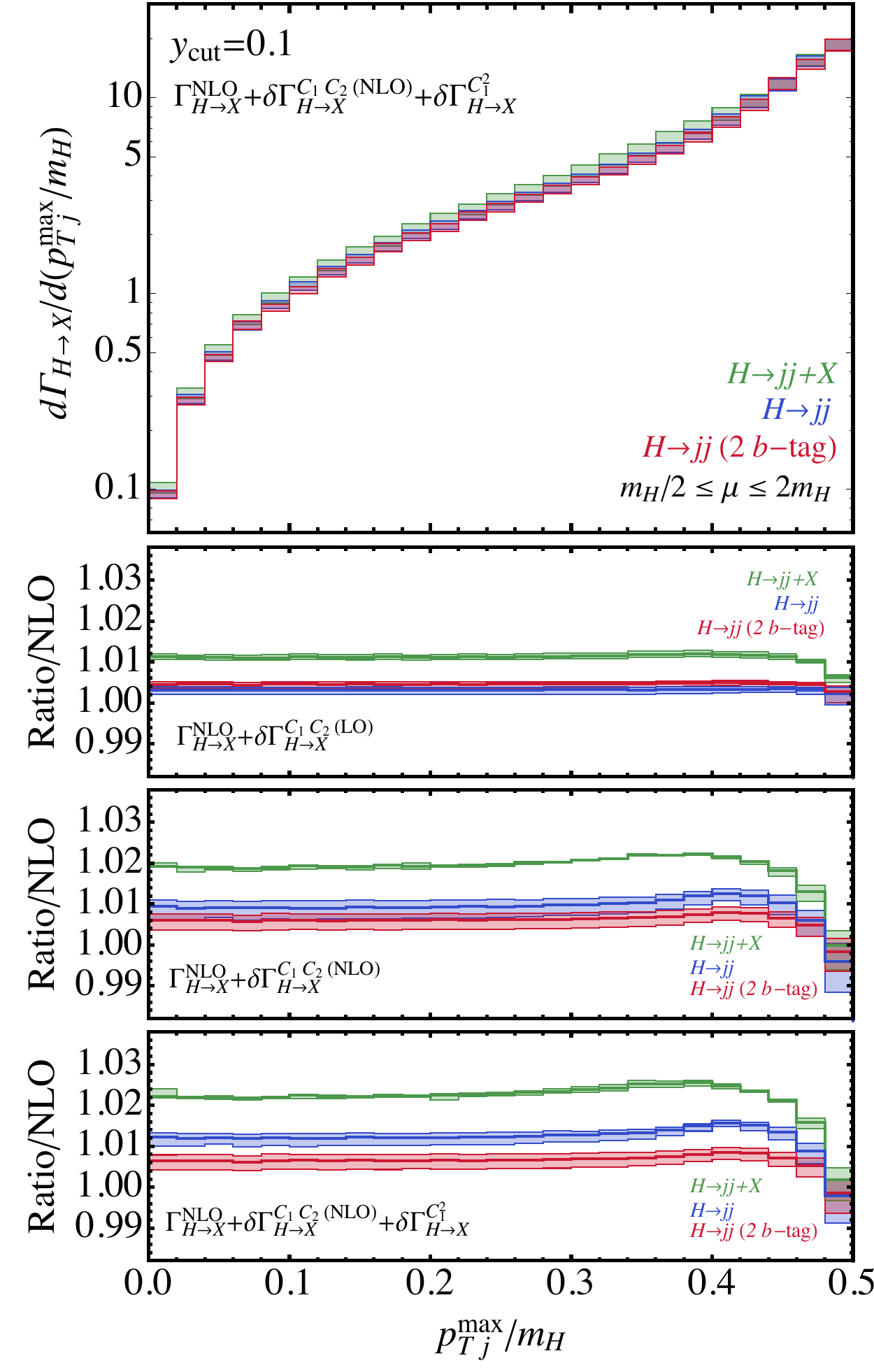}
\caption{Differential predictions for the (fractional) transverse momentum (with respect to a fictitious collision axis $\pm \hat{z}$) of the leading jet, clustered with $y_{\rm{cut}}  = 0.1$. 
 Results are presented for a variety of different requirements on the final state jets, either no restrictions $H\rightarrow jj + X$ (green), requiring exactly two jets $H\rightarrow jj$ (blue) 
 or exactly two $b$-tagged jets $H\rightarrow b\overline{b}$ (red). The lower panels present the impact of the mixed EFT-$y_b$ piece at $\mathcal{O}(\alpha_s^2)$, $\mathcal{O}(\alpha_s^3)$, and
 the combined EFT-$y_b$ and EFT$^2$ piece at $\mathcal{O}(\alpha_s^3)$, with respect to the NLO $y_b^2$ prediction. }
 \label{fig:ptj_bb}
\end{center}
\end{figure}

In fig.~\ref{fig:ptj_bb} we present the transverse momentum of the leading jet with respect to the fictitious collision axis, rescaled by the Higgs mass $m_H$. The uppermost part of the figure presents differential predictions for 
the total NLO partial width combined with the mixed EFT-$y_b$ pieces at NLO and the EFT$^2$ pieces. The lower panels show the relative impact of these new pieces compared to the basic NLO prediction. 
We present results for fully inclusive predictions $H\rightarrow jj + X$ (green),  requiring exactly two jets $H\rightarrow jj$ (blue) and finally requiring exactly two $b$-tagged jets (red). The upper most ratio-plot compares the $\mathcal{O}(\alpha_s^2)$ mixed EFT$-y_b$ to the NLO $y_b^2$ baseline. We note that in each ratio we divide by the NLO evaluated at the same scale as the numerator, and impose the final state phase-space cuts on the numerator and denominator in both instances.  In all three jet configurations we observe a reasonably flat correction, the size of which is sensitive to the 
jet requirements. At this order, the dominance of the 3-body phase-space topology can be seen by the dramatic reduction as the inclusive phase-space is reduced to the 2-jet topology. The $\mathcal{O}(\alpha_s^2)$ mixed EFT$-y_b$ terms are reduced from around a 1\% effect inclusively to 0.3\% when two-jets are mandated. Further $b$-tagging results in very little difference with respect to the two-jet predictions. 

The middle ratio panel presents the NLO predictions for the mixed EFT$-y_b$, (again with respect to the NLO $y_b^2$ pieces with equivalent cuts applied). Inclusively we see the impact of the large $K$ factor for these predictions, especially in the lower $p_T$ region, which has increased by around a factor of $2$ to $2\%$. However, at NLO the mixed EFT$-y_b$ are even more sensitive to the jet requirements, after imposing the two-jet requirements we see that these pieces are reduced to around $1\%$ of the NLO prediction. Demanding two $b$ jets further suppresses the impact of the NLO mixed EFT-$y_b$ pieces, by an additional factor of two such that 
they contribute around 0.5\% level across the phase-space. The bottom panel presents the combined NLO mixed EFT$-y_b$ piece and EFT$^2$ contribution. The inclusion of the EFT$^2$ piece increases the inclusive and $H\rightarrow jj$ pieces by around $0.3\%$, but is significantly suppressed by demanding two $b$-jets. This is anticipated since here the phase-space strongly favors the quasi-collinear limit arising from gluon splitting which rarely results in two $b$-tagged jets in the Higgs rest frame. 

It is also interesting to compare the impact of these new pieces to that of the remaining $\mathcal{O}(\alpha_s^3)$ contributions. Such a comparison is made more difficult since here we have retained the mass of the bottom quark in full, whereas in ref.~\cite{Mondini:2019gid} $m_b = 0$ kinematically. In lieu of a more detailed phenomenological study, fig.7 in ref.~\cite{Mondini:2019gid} can be inspected to provide some insight. The N3LO prediction for the $y_b^2$ pieces for this observable (for the $H\rightarrow jj$ selection criteria) is around 5\% across the bulk of phase-space (with larger impacts in the $p_T \sim m_H/2$ region). The equivalent results from our calculation here correspond to the blue curve on the bottom panel, from which we learn that the EFT component of the combined $\mathcal{O}(\alpha_s^3)$ coefficient is about 20\%.  

\begin{figure}
\begin{center}
\includegraphics[width=8cm]{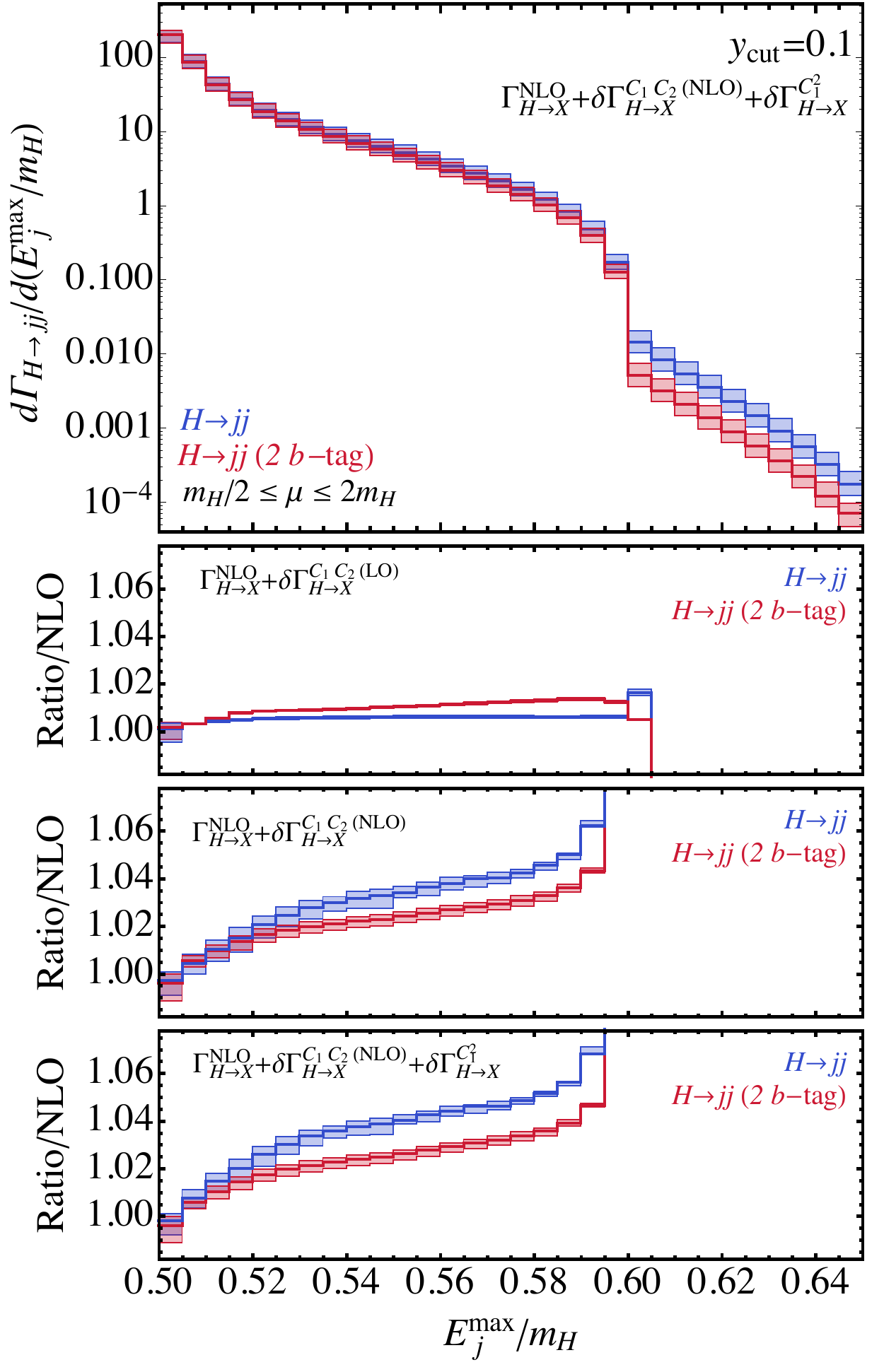}
\caption{Differential predictions for the ($m_H$-scaled) energy of the leading jet, in 2-jet events defined with $y_{\rm{cut}}  = 0.1$.
 Results are presented for a variety of different requirements on the final state jets, either requiring exactly two jets $H\rightarrow jj$ (blue) 
 or exactly two $b$-tagged jets (red). The lower panels present the impact of the mixed EFT-$y_b$ piece at $\mathcal{O}(\alpha_s^2)$, $\mathcal{O}(\alpha_s^3)$, and
 the combined EFT-$y_b$ and EFT$^2$ piece at $\mathcal{O}(\alpha_s^3)$, with respect to the NLO $y_b^2$ prediction. }
 \label{fig:Emax_bb}
\end{center}
\end{figure}

In fig.~\ref{fig:Emax_bb} we study the energy of the leading jet (rescaled by the Higgs mass) for two-jet events. This is a different type of observable than the $p_T$ studied previously, as it is a delta function at LO (where each jet has an energy of $m_H/2$). As such it is more sensitive to higher order corrections since the bulk of the phase-space is one order lower in perturbation theory. We observe a similar trend to the $p_T$ distribution studied previously. The LO mixed EFT$-y_b$ is strongly suppressed by the application of the two-jet requirement, and is a sub-percentage effect across the phase-space. The higher order corrections to the mixed EFT-$y_b$ term opens up a more dynamic three-parton and four-parton phase-space and results in relatively large corrections (middle ratio panel). Without specifically requiring $b$-tagging the
contribution to the bulk of the phase-space is around 3-4\%, which is reduced to around 2-3\% when the $b$-tagging algorithm is applied. The $H\rightarrow jj$ rate is increased by around 0.5-1\% by the inclusion of the EFT$^2$ contribution, but this again is significantly reduced by the requirement of two $b$-tags. Finally comparison with fig. 8, in ref.~\cite{Mondini:2019gid} shows that the EFT initiated effects are rather small compared to the total impact of the other $\alpha_s^3$ pieces with bulk corrections around 30-50\% for the combined $\alpha_s^2$ and $\alpha_s^3$ coefficient. We note however, that although these residual EFT effects are small (~10\% of the total $\alpha_s^3$ contribution), their magnitude is comparable to, or larger than, the scale uncertainty at this order. The far tail in this distribution is only directly accessible by the four-parton contribution (which exists in the NLO correction to the mixed EFT-$y_b$ pieces), and a comparison with the NLO partial width is not valid. 

Summarizing, in this section we have presented a short study of the EFT initiated pieces at $\mathcal{O}(\alpha_s^3)$. We found that the impact of these pieces is sensitive to the nature of the final state selection cuts. This is hardly surprising, given that this contribution arises form an interference of two disparate phase-spaces. The pure-EFT diagrams are largest when the $g\rightarrow b\overline{b}$ splitting is quasi-collinear which suppresses the two $b$-jet rate significantly. Therefore the impact of these pieces varies from a few precent (and hence crucial for future precision studies), to sub-percent (and less relevant to phenomenology) depending on the specific final state selection criteria employed. It is difficult to speculate on LHC applications given that we have used the Durham jet algorithm in the Higgs rest-frame, but its seems reasonable to postulate that analysis which require two resolved $b$-jets will have very small impacts from the EFT initiated pieces, whereas those that include boosted jets include more of the $g\rightarrow b\overline{b}$ splitting pieces and are more sensitive to the EFT pieces. However, the uncertainties with this type of analysis are rather large and unlikely to be at the percent level in the near future. In the longer term, final studies at the HL-LHC and proposed FC analyses aiming for precent precision accuracy  should carefully model these pieces.

\section{Results for {\hcc}}
\label{sec:hccres}

\subsection{Inclusive results} 

In this section we present an analysis of the \hcc decay.  We use a charm pole mass of $m_c = 1.67$ GeV which is used in the kinematics and propagators, 
in the Yukawa coupling we use $m_c^{\overline{{\rm{MS}}}} = \{0.65, 0.61,0.58\}$ GeV at the scales $\{0.5,1,2\} \times m_H$ respectively. All other parameters
are the same as defined in our computations of \hbb. 

\begin{table}
\begin{center}
\def\arraystretch{1.7}
\begin{tabular}{|c |c || c c c |} 
\hline
& Partial Width [MeV] & $\mu = m_H/2$ & $\mu = m_H$ & $\mu = 2 m_H$ \\
\hline
$\mathcal{O}(\as^0)$ & $\Gamma^{(0)}_{H\rightarrow c\overline{c}}$ & 0.1047 &  0.0930&  0.0836 \\
$\mathcal{O}(\as^1)$ & $\Gamma^{(1)}_{H\rightarrow c\overline{c}}$ & 0.0122 & 0.0190  & 0.0231 \\
\hline 
\hline
$\mathcal{O}(\as^2)$ &  $\delta\Gamma_{H\rightarrow c\overline{c}}^{C_3^2}$, [$y_c^2$]   &    -0.0005  &  0.0035 &  0.0069 \\
& $\delta\Gamma_{H\rightarrow c\overline{c}}^{C_1C_3}$ [EFT-$y_c$]  & 0.0036   & 0.0030 &  0.0026  \\
& $\delta\Gamma_{H\rightarrow c\overline{c}}^{C_2C_3}$  [$y_b y_c$] & 0.0001  & 0.00007 &  0.00005  \\
& $\delta\Gamma_{H\rightarrow c\overline{c}}^{C_2^2}$  [$y_b^2$] & $\sim 10^{-6}$  & $\sim 10^{-6}$  & $\sim 10^{-6}$    \\
\hline 
\hline
$\mathcal{O}(\as^3)$ &  $\delta\Gamma_{H\rightarrow c\overline{c}}^{C_3^2}$  [$y_c^2$]    & -0.0008  &  0.0002  &  0.0017  \\
&$\delta\Gamma_{H\rightarrow c\overline{c}}^{C_1C_3}$   [EFT-$y_c$]  & 0.0020 & 0.0023   &  0.0023   \\
& $\delta\Gamma_{H\rightarrow c\overline{c}}^{C_1^2}$  [EFT$^2$] & 0.0154  & 0.0112 & 0.0084  \\
\hline 
\end{tabular}
\caption{Inclusive partial width results for \hcc at LO, and higher order coefficients for NLO $y_c^2$ contributions, and the top induced pieces at $\mathcal{O}(\alpha_s^2)$ and $\mathcal{O}(\alpha_s^3)$.} 
\label{tab:incytyc}
\end{center}
\end{table}

In Table~\ref{tab:incytyc} we present a summary of the inclusive contributions to final states involving charm-quarks. From a comparison with the equivalent table for bottom quarks (Table~\ref{tab:incybyt}) we see that the relative impact for the EFT pieces is much larger for charm quarks. Taking $\mu=m_H$ as a reference point we see that for the charm decays the EFT-$y_c$ piece is the same size as the $y_c^2$ correction at $\mathcal{O}(\alpha_s^2)$, and a factor of 10 bigger at $\mathcal{O}(\alpha_s^3)$. For comparison the same ratios for \hbb are approximately 33\% and three.  The relative enhancements come from two sources. 
For a given quark species $f$, the overall scaling of the mixed-EFT terms to the LO behaves like, 
\begin{eqnarray}
\frac{y_f  m_f}{y_f^2} \propto  \frac{m_f}{m_f^{\msbar}(m_H)} 
\end{eqnarray}
which is equal to around $2.5$ for charm quarks and $1.67$ for bottoms. Thus the mixed terms are relatively more important for the charm quarks. Secondly, the smaller charm mass causes a larger quasi-collinear enhancement of the $g\rightarrow c \overline{c}$ splitting (compared to $g\rightarrow b \bb$) which increases the importance of the EFT amplitude. However, there is a difference with respect to the bottom quark, for charm's there is a destructive interference between the two-body phase-space and the three-body phase-space at LO.  This destructive interference is also present at NLO, but the inclusion of additional positive contributions from the four-body phase-space conspire to produce a rather sizable $K$-factor $(1.5-2)$ in going to NLO in the mixed EFT-$y_c$ pieces. 

A new feature of the \hcc partial widths is the $y_b y_c$ mixed term and $y_b^2$ term which appear at $\mathcal{O}(\alpha_s^2)$. As discussed in section~\ref{sec:calc} there is an ambiguity as to whether one includes these contributions in either the \hbb or \hcc partial widths. In our setup it makes most sense to include them in the discussion of \hcc, since $m_c=0$ in our \hbb and accordingly these terms either vanish or are IR-unsafe in isolation. Either way, it is clear from the results in Table~\ref{tab:incytyc} that these pieces are rather small and unless targeted with a particularly exotic experimental selection requirement, are of limited phenomenological relevance. The $y_b^2$ term, corresponding to the four-quark amplitudes interfering at tree-level is very small ($\sim $ 1 eV), the mixed $y_by_c$ is dominated by the 2-body term. Given the smallness at $\mathcal{O}(\alpha_s^2)$, and its technical complexity, we do not study these terms at $\mathcal{O}(\alpha_s^3)$. Table~\ref{tab:incytyc} also highlights the large impact of the EFT$^2$ contributions to the inclusive width. At around 10\% of the LO partial width, these pieces dominate all corrections beyond NLO in the expansion.  Their increase in relative importance is because they do not suffer any suppression when moving from bottom to charm quarks in the final state, and are in fact enhanced by the increased effect of the quasi-collinear splitting due to smaller charm quark mass ($\log(m_b^2/m_c^2) \sim 2$). For the \hbb decay we observed a strong sensitivity to jet selection criteria for these pieces, therefore in the next subsection we will turn our attention to a differential analysis. 

\subsection{Differential predictions for EFT-induced contributions to \hcc}

\begin{figure}
\begin{center}
\includegraphics[width=8cm]{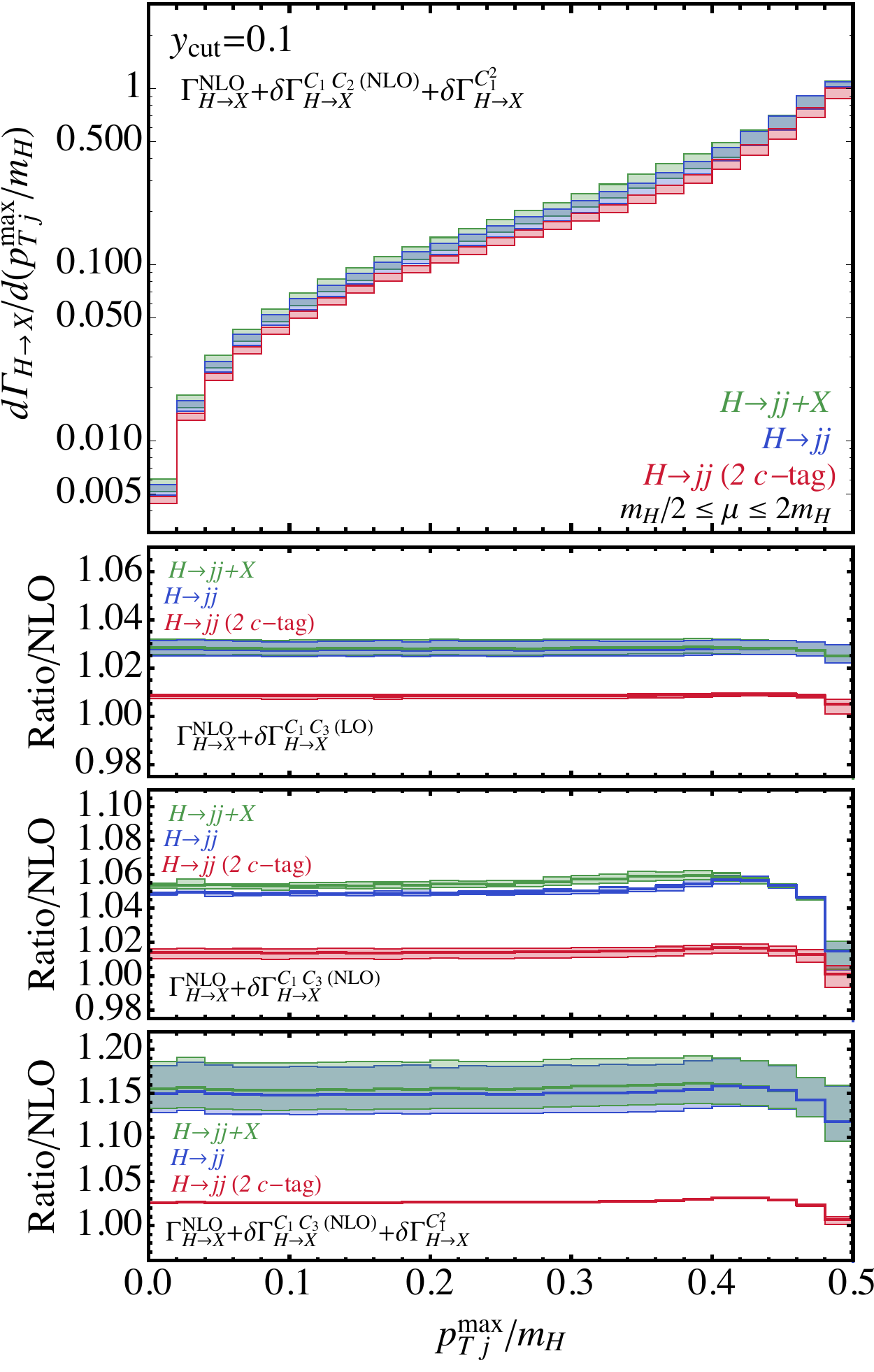}
\caption{Differential predictions for the (fractional) transverse momentum (with respect to a fictitious collision axis $\pm \hat{z}$) of the leading jet, clustered with $y_{\rm{cut}}  = 0.1$. 
 Results are presented for a variety of different requirements on the final state jets, either no restrictions $H\rightarrow jj + X$ (green), requiring exactly two jets $H\rightarrow jj$ (blue) 
 or exactly two $c$-tagged jets (red). The lower panels present the impact of the mixed EFT-$y_c$ piece at $\mathcal{O}(\alpha_s^2)$, $\mathcal{O}(\alpha_s^3)$, and
 the combined EFT-$y_c$ and EFT$^2$ piece at $\mathcal{O}(\alpha_s^3)$, with respect to the NLO $y_c^2$ prediction evaluated with the same cuts. }
 \label{fig:ptj_cc}
\end{center}
\end{figure}
\begin{figure}
\begin{center}
\includegraphics[width=8cm]{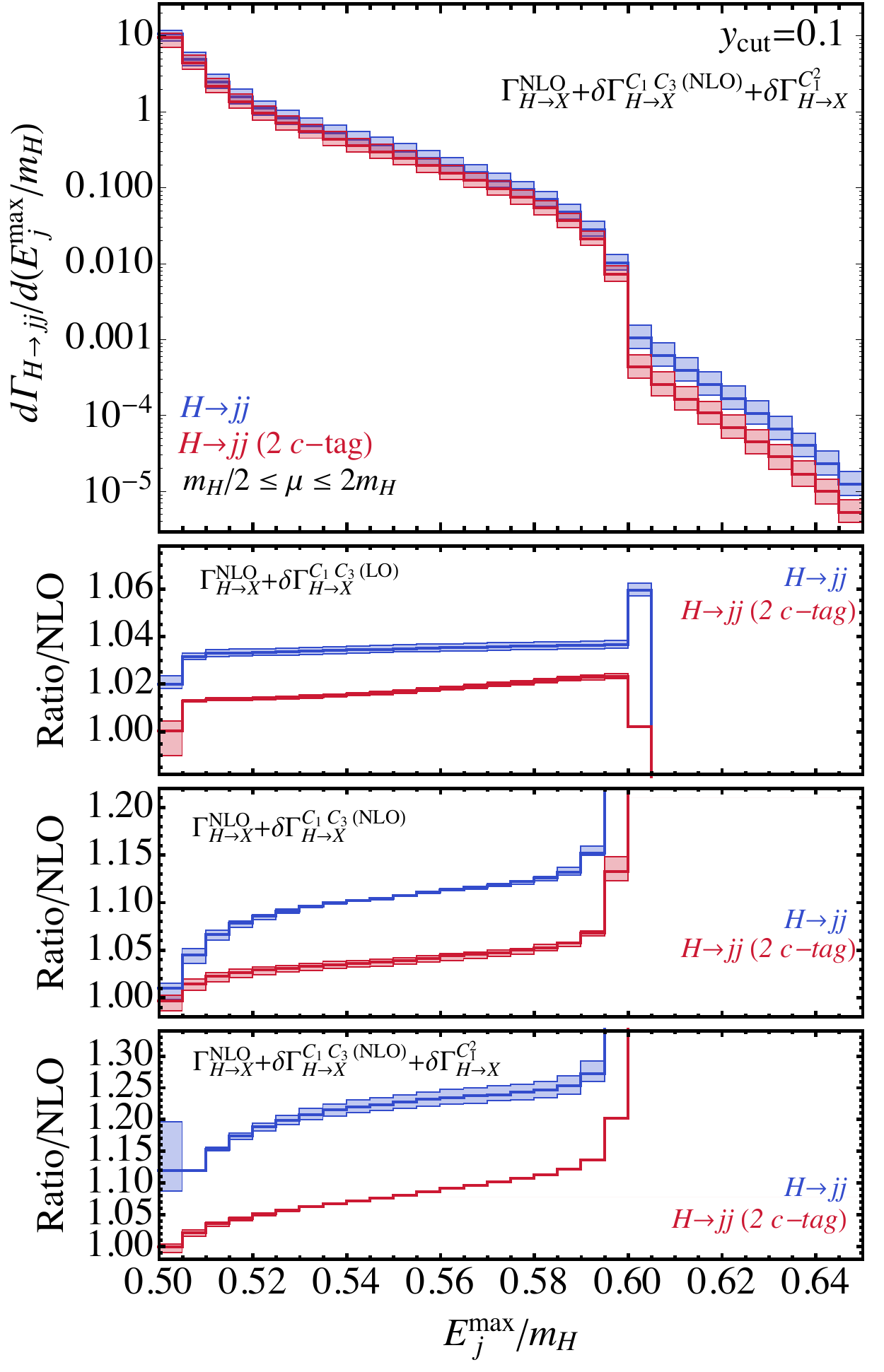}
\caption{Differential predictions for the ($m_H$-scaled) energy of the leading jet in 2-jet events (in \hcc decays) defined with $y_{\rm{cut}}  = 0.1$.
 Results are presented for a variety of different requirements on the final-state jets, either requiring exactly two jets $H\rightarrow jj$ (blue) 
 or exactly two $c$-tagged jets (red). The lower panels present the impact of the mixed EFT-$y_c$ piece at $\mathcal{O}(\alpha_s^2)$ and $\mathcal{O}(\alpha_s^3)$, and
 the combined EFT-$y_c$ and EFT$^2$ pieces at $\mathcal{O}(\alpha_s^3)$, with respect to the NLO $y_c^2$ prediction. }
 \label{fig:Emax_cc}
\end{center}
\end{figure}

In this section we compute differential predictions for \hcc, again focussing on the pieces arising from the EFT contributions at NLO. As in the analysis for \hbb we use the Durham jet algorithm with $y_{\rm{cut}}  = 0.1$, we compute the transverse momentum and energy (rescaled by the Higgs mass)  for the leading jet in 2-jet and 2 $c-$tagged jet selection cuts. Our results are shown for the transverse momentum in fig.~\ref{fig:ptj_cc}, and energy in fig.~\ref{fig:Emax_cc}. 
As before the upper panel shows the differential distribution, and the lower panels present various ratios to the respective NLO predictions (with the selection cuts applied). 
Focusing first on the transverse momentum,
inspection of the lower panels shows that the mixed EFT-$y_c$ terms at NLO correspond to around a 5\% correction across the bulk of the distribution, but are again significantly reduced by demanding exactly two $c$-tagged jets in the final state. The bottom panel highlights the huge impact of the EFT$^2$ contribution inclusively, around 15\%.  However, once again, the requirement that the two charm quarks reside in different tagged jets significantly damps the relative importance of the contribution. The total EFT sensitive pieces (mixed + squared) contribute around the 2-3\% of the NLO if two $c$-jets are required. A similar story is shown in the energy distribution, across the bulk of the phase-space only the 3- and 4-body phase-space configurations contribute, resulting in a larger relative impact of the EFT pieces. Again, application of a charm tagging algorithm significantly reduces these contributions, by around a factor of four. 

\subsection{Impact on LHC (and FC) studies of \hcc}

Finally, we conclude this section with a brief comment  regarding the impact of the EFT induced pieces on \hcc measurements at the LHC. Summing the contributions in Table~\ref{tab:incytyc} and rescaling $y_c = \kappa_c y_c$ and $C_1^0 = \kappa_g C_1^0$ (i.e. each $\kappa_X$ picks out the pure-charm and pure-EFT amplitudes, and $\kappa_X=1$ in the SM) we can write
\begin{eqnarray}
\Gamma^{{\rm{N3LO}}}_{H\rightarrow c\overline{c} +X}[\mu=m_H] =  0.0967 \left(\kappa_c^2 + 0.055 \times \kappa_c \kappa_g + 0.116 \times \kappa_g^2 \right) + \mathcal{O}(\alpha_s^4) \quad {\rm{MeV}}
\end{eqnarray}
Thus there is room for a significant mismeasurement $(\sim 15\%)$ of $\kappa_c$ if $\kappa_g = 0$  is enforced in the above equation. One should therefore ensure that these pieces are adequately modeled if an extraction of $y_c$ is attempted at a level of 20\% or better. As a rule of thumb analyses which are sensitive to $g\rightarrow c\overline{c}$ splittings will have a large impact from $\kappa_g$ pieces. For example, requiring two isolated charm-tagged jets will likely reduce  the contributions down to the level of a few percent, whereas analyses that allow QCD radiation to fall into the same jet (e.g. boosted) searches, will be more exposed to this type of correction. Interestingly, both types of analysis are currently employed in the experimental analyses~\cite{Sirunyan:2019qia}. We leave a more detailed LHC phenomenological study to a future publication. 

\section{Conclusions} 
\label{sec:conc}
In this paper we have presented a calculation of the EFT-sensitive pieces arising in $H\rightarrow q\overline{q}$ decays at $\mathcal{O}(\alpha_s^3)$. We have retained the mass of the final-state quarks  throughout 
our calculation, and worked in an EFT in which the top quark is integrated out of the theory. 
This calculation required the computation of the two-loop amplitudes in the EFT interfered with the tree-level $H\rightarrow q\overline{q}$ amplitude and
the one-loop corrections in the EFT interfered with the one-loop corrections to $H\rightarrow q\overline{q}$ (2-body phase-space contributions). The three-body 
phase-space consisted of a one-loop EFT (or $y_q$) $H\rightarrow q\overline{q}g$ amplitude interfered with the $y_q$ (or EFT) tree-level amplitude. Finally, there are four-body terms which are tree-level four-parton amplitudes in the EFT interfered with those proportional to $y_q$. 

These EFT-induced pieces are computationally interesting, since the 
$\mathcal{O}(\alpha_s^2)$ amplitudes are UV divergent, but IR-finite. Our computation therefore bears some hall-marks of an NLO computation and some of an NNLO computation in QCD. 
The three contributing phase-space regions act like double-virtual, real-virtual, and double-real terms in a traditional NNLO calculation, but they can be regulated using a NLO dipole subtraction setup. 
The UV renormalization at this order is intricate, especially given the mixed renormalization scheme employed, in which the Yukawa coupling is renormalized in the $\msbar$ scheme while the bottom quark mass is renormalized in the on-shell scheme. 
As a result, the amplitudes scale like $y_q^{\msbar} m_q^{\rm{OS}}$. 

We implemented our results into a Monte Carlo code (based upon MCFM) capable of simulating the full kinematics of the decay products. We subsequently used this code to produce differential predictions for the $H\rightarrow b\overline{b}$ decay in the Higgs boson rest frame, using the Durham jet algorithm. We found that the impact of the EFT-initiated pieces is very sensitive to the specific nature of the final-state phase-space selection criteria. Since the 2-body phase-space is suppressed relative to the 3-body one (at $\mathcal{O}(\alpha_s^2)$ where a clear distinction between the two can be made), requiring exactly two jets suppresses the impact of the EFT pieces. Further suppression occurs when two $b$-tagged jets are demanded, which can be also be traced back to the 3-body topology, since this matrix element is largest when the $g\rightarrow b\overline{b}$ splitting is quasi-collinear and the typical contribution is a single $b$-jet (containing the $b$ and $\overline{b}$ pair) and a gluon jet. These effects are even more pronounced for the final-state charm quark, where due to the fact that $y_c < y_b$ the EFT$^2$ pieces are relatively enhanced with respect to the LO. Due to quasi-collinear logarithms, the mixed EFT terms are also more important for the \hcc channel.

The EFT-initiated pieces spoil the relationship $\Gamma_{H\rightarrow q\overline{q}} \propto y_q^2$ and thus could pollute an extraction of the Higgs-quark coupling at colliders. Our results can be used to quantify the (QCD) effect of non-$y_q^2$ pieces, and their role in collider observables. The effects are maximally around a few percent (bottom) or 10-15\% (charm), but can be significantly suppressed by the selection criteria. Therefore, any analysis at the HL-LHC, or especially future lepton colliders, should endeavor to model these pieces given the fiducial specifics of the analysis. Finally we note that a full phenomenological study at the LHC and FC of $H\rightarrow q\overline{q}$ including the full $\mathcal{O}(\alpha_s^3)$ prediction and EW corrections (which also induce a $y_t$ dependence) is extremely motivated. We leave this analysis to a future study. 

\acknowledgments

The authors are supported by a National Science Foundation CAREER award number PHY-1652066.
U.S. is additionally supported by the National Science Foundation award number PHY-1719690.
Support provided by the Center for Computational Research at the University at Buffalo.

\appendix

\section{Renormalization coefficients }
\label{sec:appRenom}

In this appendix we provide a list of the renormalization coefficients needed in our calculation. 
The gluon wavefunction renormalization coefficient is given by 
\begin{align} \label{eqZ31l}
Z_A^\os =1 -\frac{\as}{4\pi} \frac{4 \, T_R}{3 \eps} \Gamma(1+\eps) \sum_{i \in n_h} \left(\frac{\mu^2}{m_i^2} \right)^{\eps} + \order{\as^2} \, ,
\end{align}
and for the strong coupling $\alpha_s$,
\begin{align} \label{eqZg1l}
Z_{\as}^{\msbar} =1 -\frac{\as}{4\pi}  \frac{1}{\eps} \left (\frac{11}{3} C_A - \frac{8}{3} T_R \, n_f \right ) + \mathcal{O}(\alpha_s^2) \, ,
\end{align}
where $n_f$ is the combined number of light flavors and heavy flavors ($n_f =5$ in our calculations). For the heavy quark wavefunction, 
\begin{align} \label{eqZ21l}
Z_Q^\os =1 -\frac{\alpha_s}{4\pi}C_f  \, \Gamma(1+\eps) \left(\frac{\mu^2}{m_Q^2} \right)^{\eps}  \frac{(3-2\eps)}{\eps \,(1-2\eps)} + \mathcal{O}(\alpha_s^2)
\end{align}
and heavy quark masses, 
\begin{align} \label{eqZm1l}
Z_{m_Q}^\os = -\frac{\alpha_s}{4\pi} C_f  \, \Gamma(1+\eps) \left(\frac{\mu^2}{m_Q^2} \right)^{\eps} \frac{(3-2\eps)}{\eps \,(1-2\eps)} + \mathcal{O}(\alpha_s^2) \, .
\end{align}
Finally, the Yukawa couplings are renormalized using
\begin{eqnarray}
Z_{y_b}^{\msbar} &=&1+\frac{\alpha_s}{4\pi} \left(-\frac{3C_f}{\eps} \right) + \left(\frac{\alpha_s}{4\pi}\right)^2 \bigg[ C_f^2 \left( \frac{9}{2 \eps^2} -\frac{3}{4\eps}  \right)  \nonumber\\
& &+ C_f \, C_A \left(\frac{11}{2 \eps^2} - \frac{97}{12\eps}  \right) + C_f n_f \left( -\frac{1}{\eps^2}+\frac{5}{6\eps}\right)  \bigg]+ \mathcal{O}(\alpha_s^3) \, .
\label{eqZm1l}
\end{eqnarray}
In order to renormalize the Wilson coefficient $C_2$, the heavy-quark mass renormalization is also needed in the $\msbar$ scheme,
\begin{eqnarray}
Z_m^{\msbar}&=&1+\frac{\alpha_s}{4\pi} \left(-\frac{3C_f}{\eps} \right) + \left(\frac{\alpha_s}{4\pi}\right)^2 \bigg[ C_f^2 \left( \frac{9}{2 \eps^2} -\frac{3}{4\eps}  \right)  \nonumber\\&&
 + C_f \, C_A \left(\frac{11}{2 \eps^2} - \frac{97}{12\eps}  \right) + C_f n_f \left( -\frac{1}{\eps^2}+\frac{5}{6\eps}\right)  \bigg] + \mathcal{O}(\alpha_s^3) \, .
\end{eqnarray}
Eqs.~\eqref{eq:c1def}$-$\eqref{eq:c2def} define the matching coefficients $C_1$ and $C_2$ in terms of $\Delta_H^{(i)}$ and $\Delta_F^{(i)}$. In this calculation for $C_1$ we need the following~\cite{Chetyrkin:1997un}, 
\begin{eqnarray}
\Delta_{H}^{{\rm{(1)}}} = \left(\frac{11}{4}-\frac{1}{6} \log\frac{\mu^2}{m_t^2}   \right) \, ,
\end{eqnarray}
and for $C_2$,
\begin{eqnarray}
\Delta_{F}^{{\rm{(2)}}} &=& \left( \frac{5}{18}-\frac{1}{3}\log \frac{\mu^2}{m_t^2} \right) \\
\Delta_{F}^{{\rm{(3)}}} &=&\frac{311}{1296}+\frac{5}{3}\zeta(3)-\frac{175}{108}\log \frac{\mu^2}{m_t^2}-\frac{29}{36}\log^2\frac{\mu^2}{m_t^2}
    +n_l\left(\frac{53}{216}+\frac{1}{18}\log^2 \frac{\mu^2}{m_t^2} \right) \, ,
\end{eqnarray}
where we note that these results are defined in the $\msbar{}$ scheme with $N_c=3$.

\section{Definition of canonical master integrals}
\label{app:MI}
\begin{figure}
  \centering
  \captionsetup[subfigure]{labelformat=empty}
  \subfloat[$\mathcal{T}_1$]{%
    \includegraphics[width=0.105\textwidth]{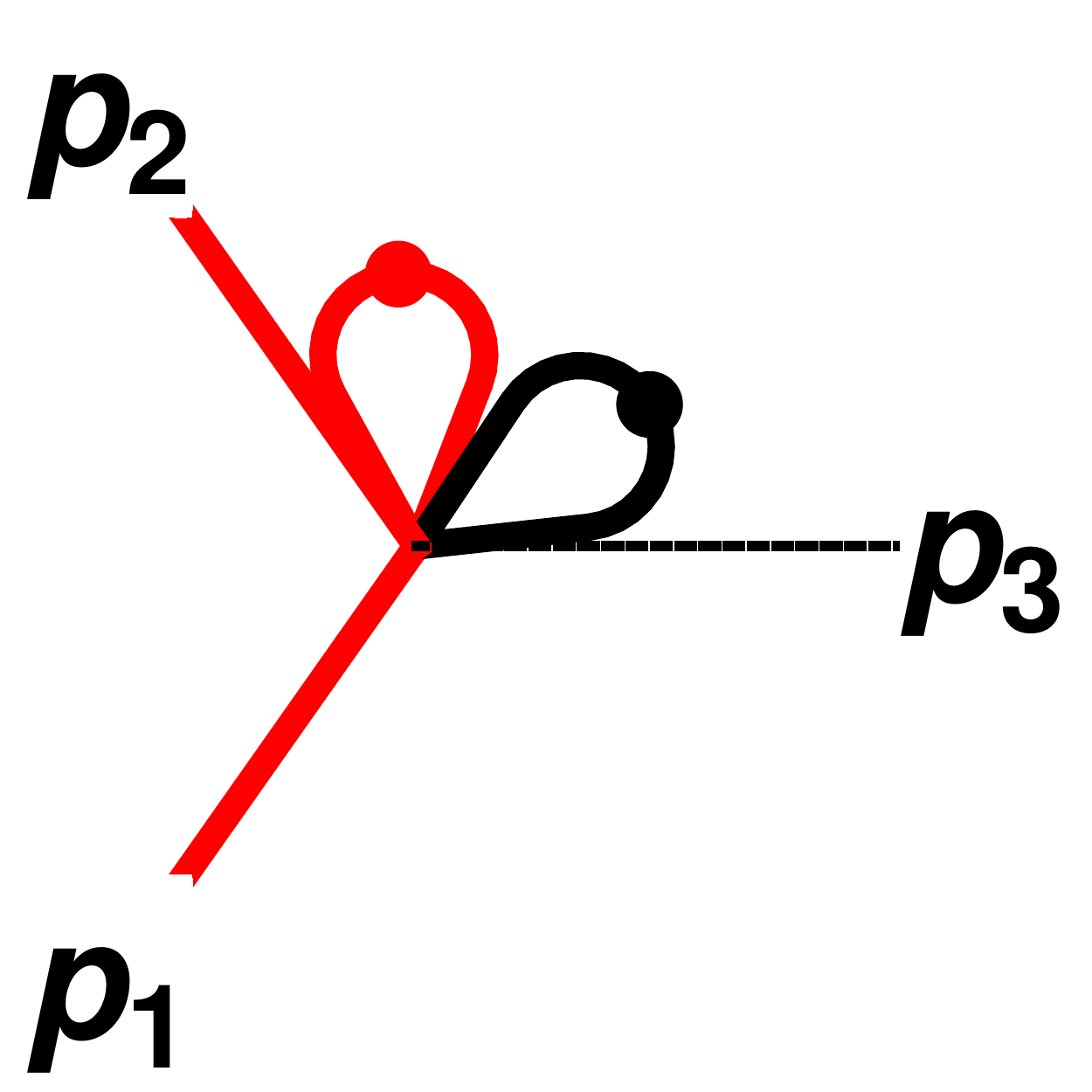}
  } \,
  \subfloat[$\mathcal{T}_2$]{%
    \includegraphics[width=0.105\textwidth]{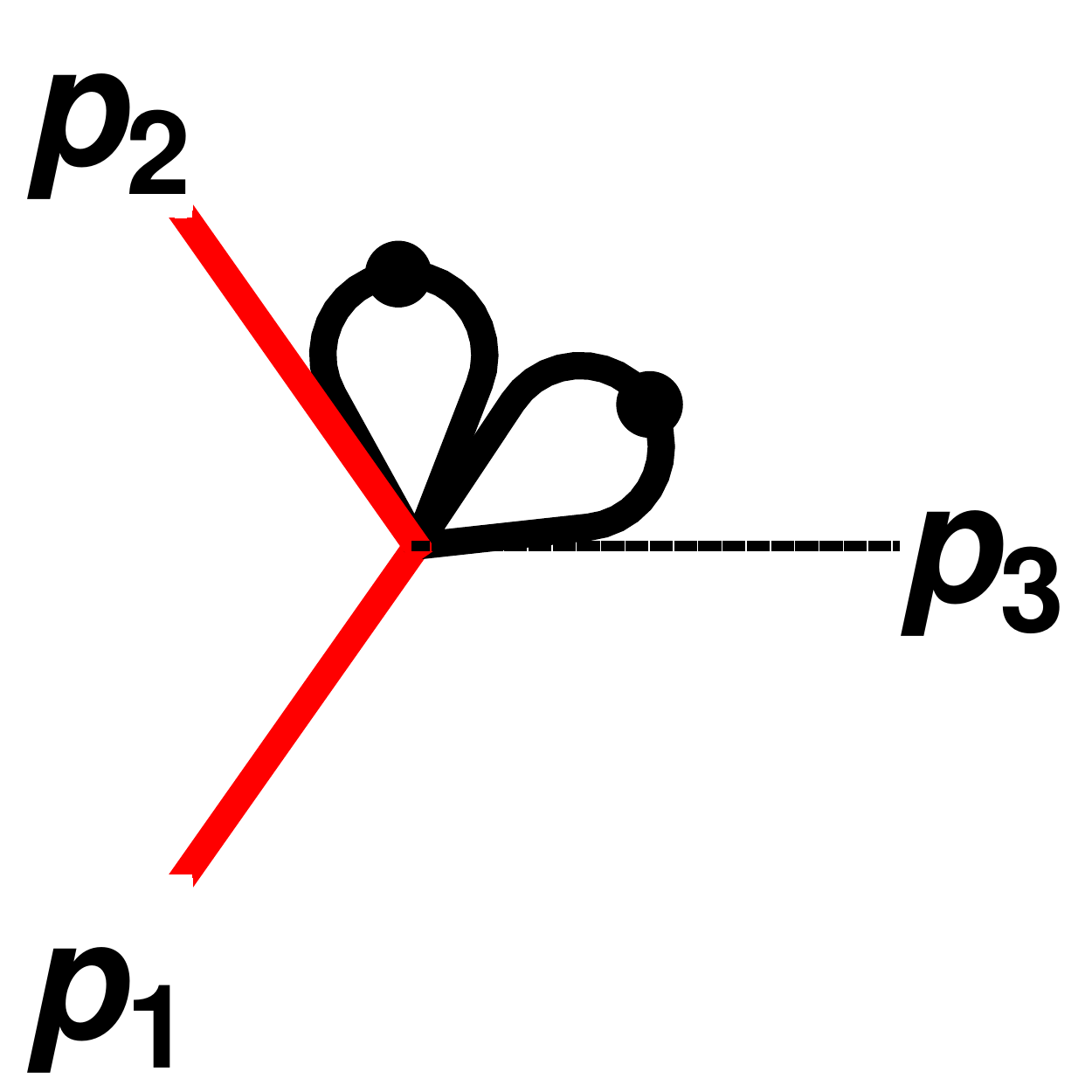}
  }\,
  \subfloat[$\mathcal{T}_3$]{%
    \includegraphics[width=0.105\textwidth]{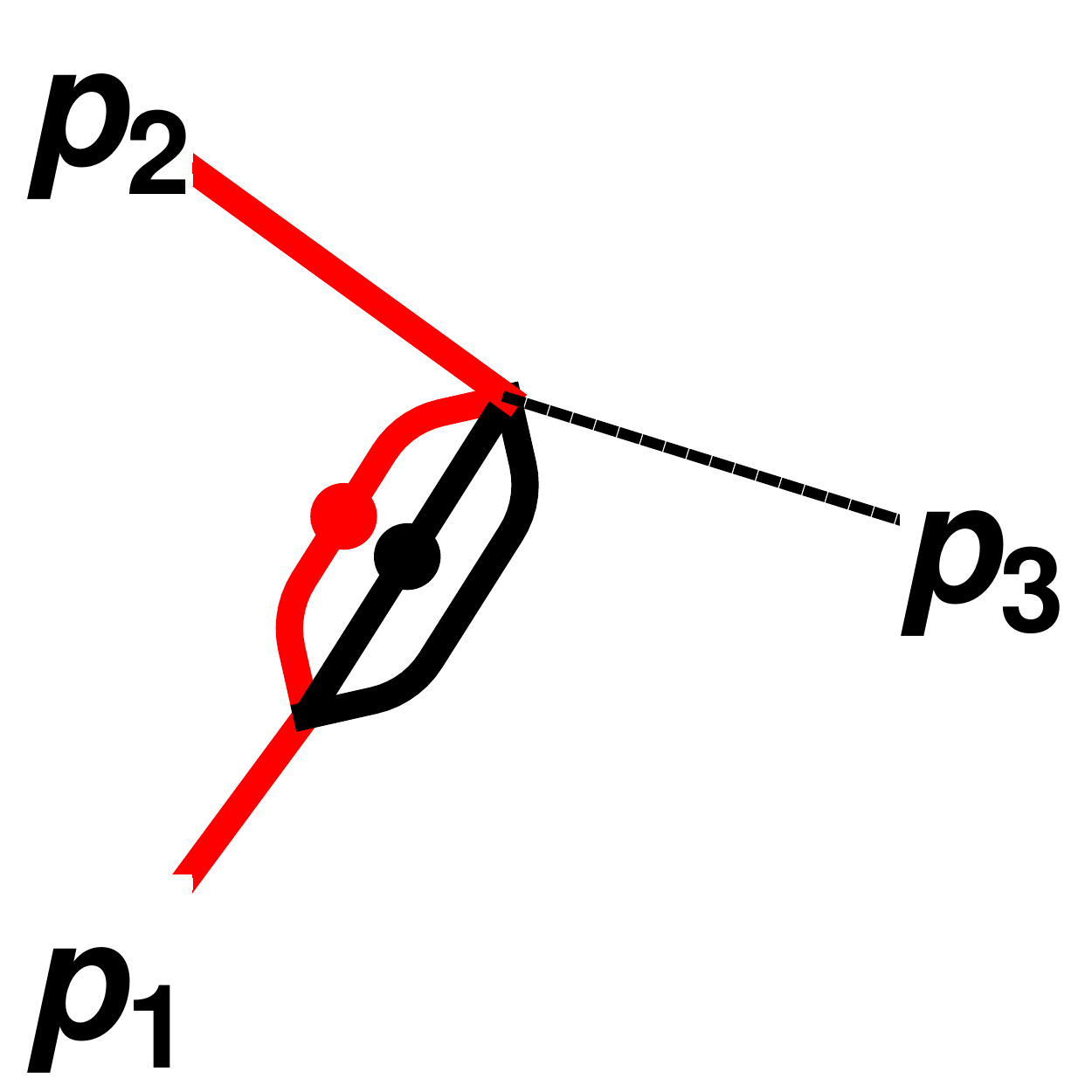}
  }\,
  \subfloat[$\mathcal{T}_4$]{%
    \includegraphics[width=0.105\textwidth]{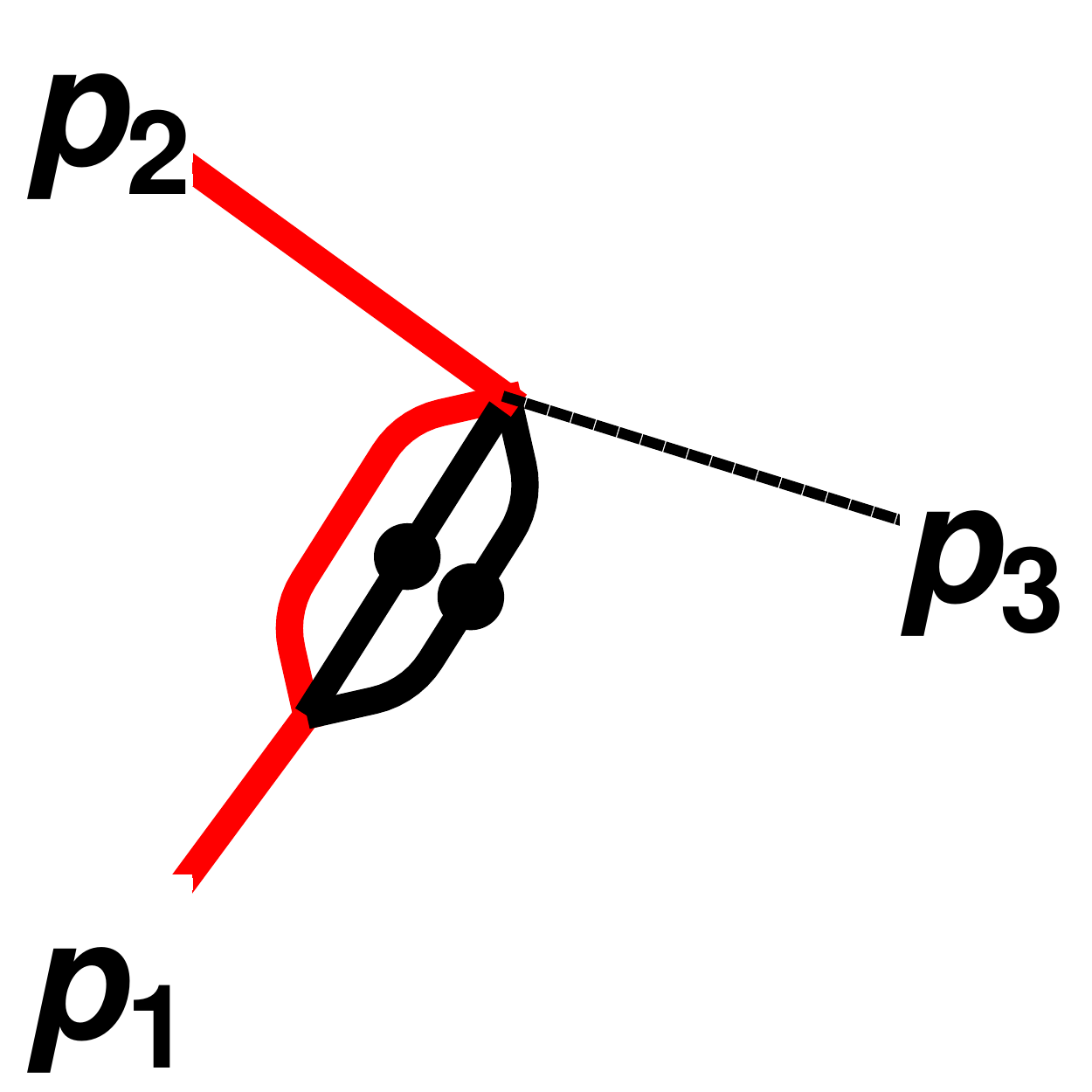}
  }\,
  \subfloat[$\mathcal{T}_5$]{%
    \includegraphics[width=0.105\textwidth]{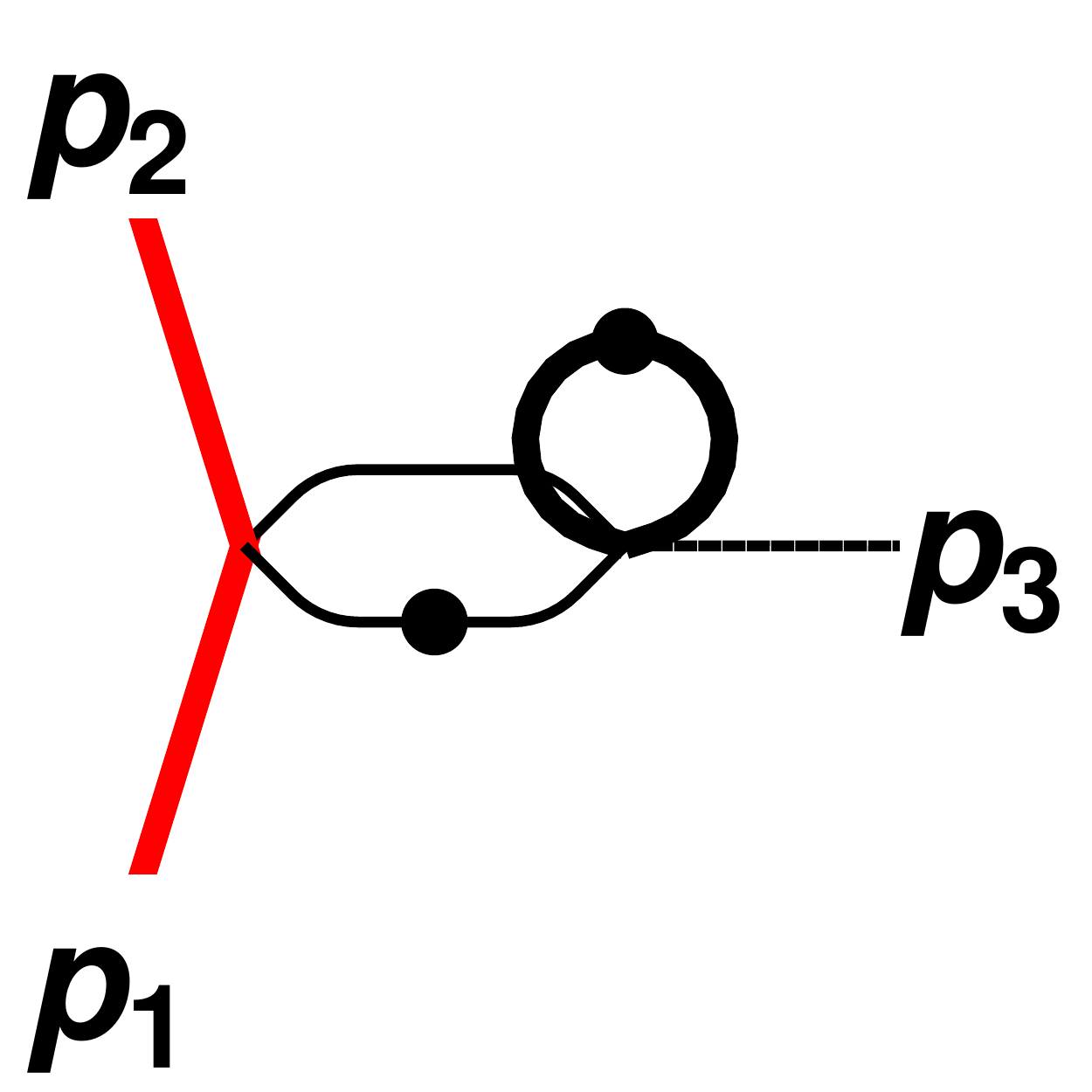}
  } \,
  \subfloat[$\mathcal{T}_6$]{%
    \includegraphics[width=0.105\textwidth]{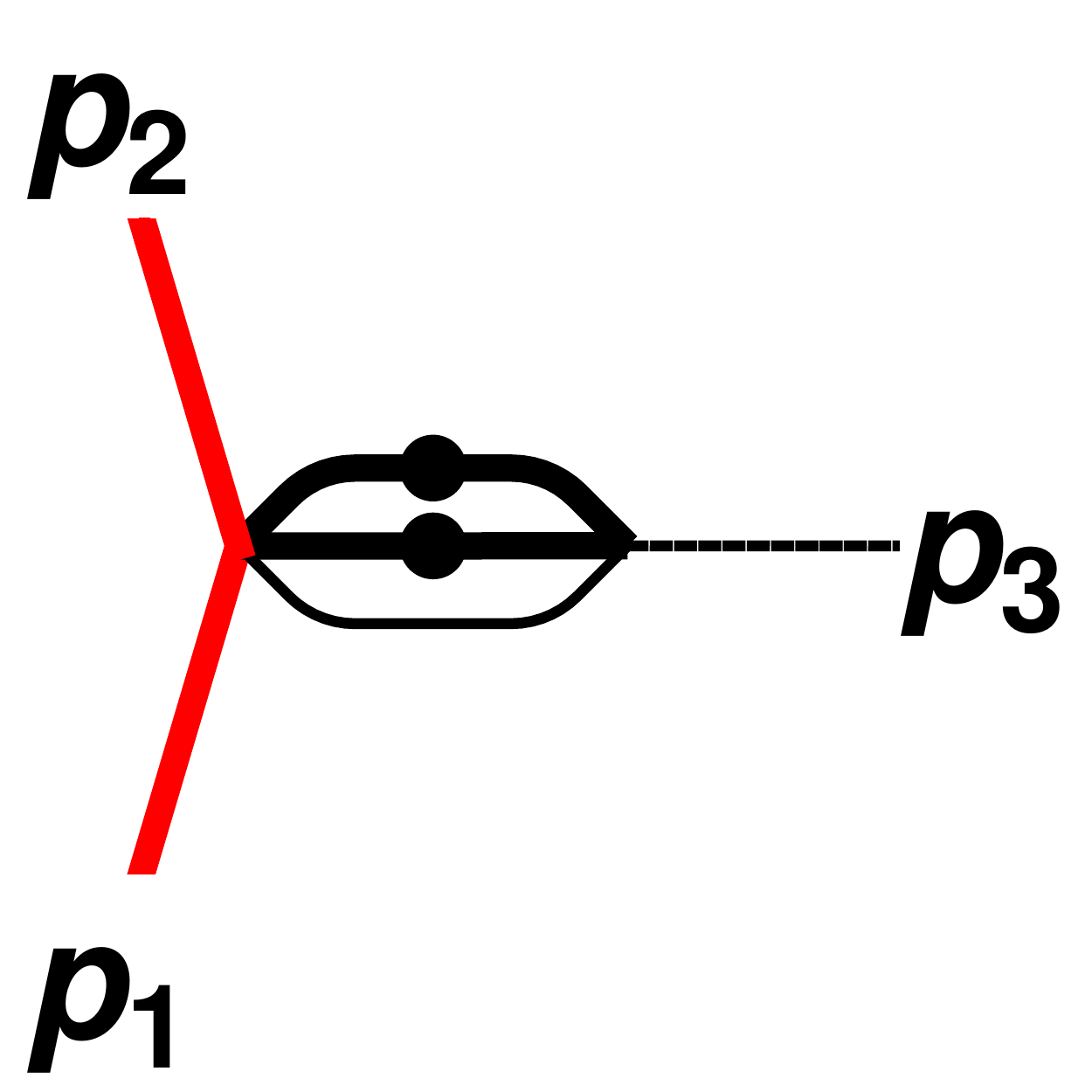}
  } \,
  \subfloat[$\mathcal{T}_7$]{%
    \includegraphics[width=0.105\textwidth]{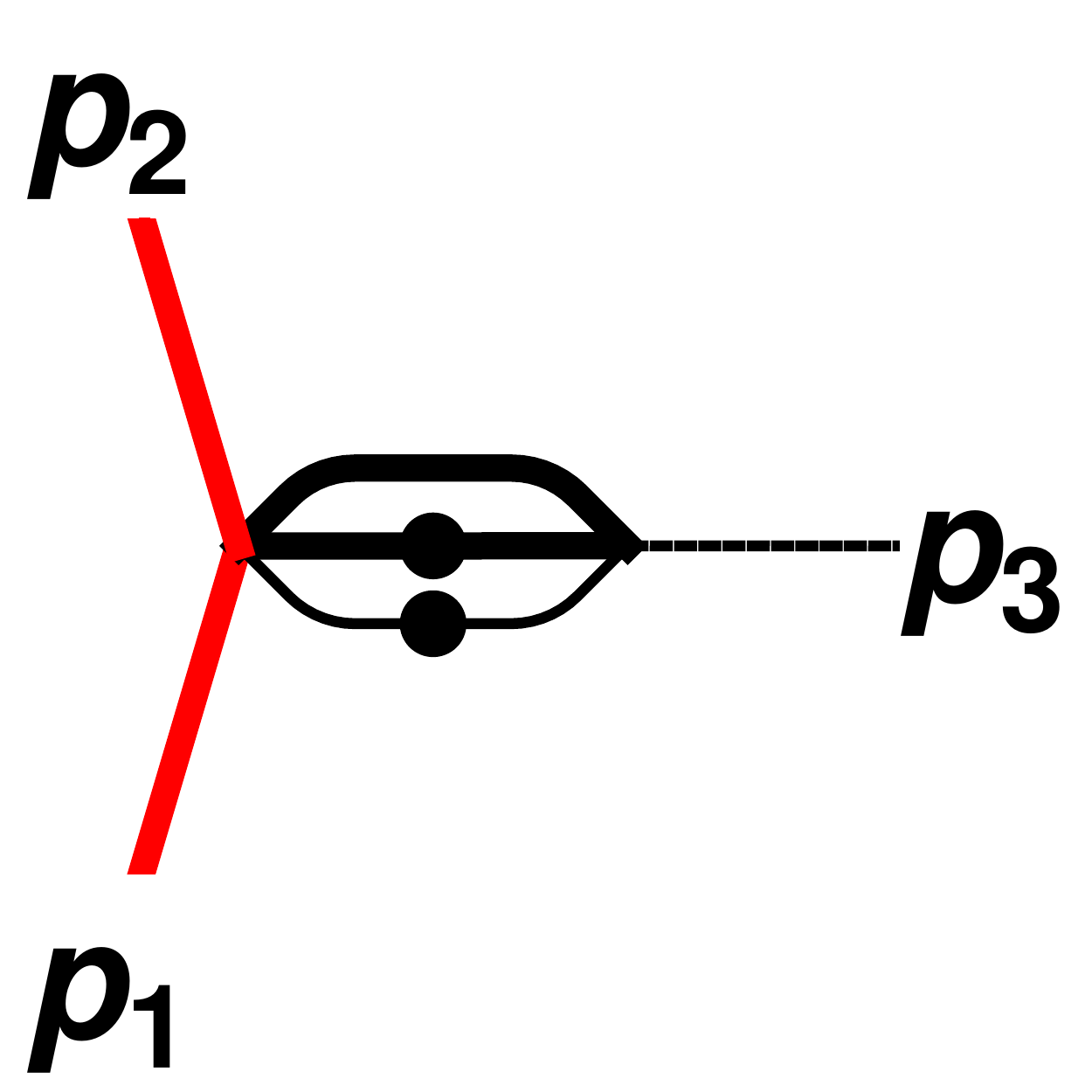}
  }\\
  \subfloat[$\mathcal{T}_8$]{%
    \includegraphics[width=0.105\textwidth]{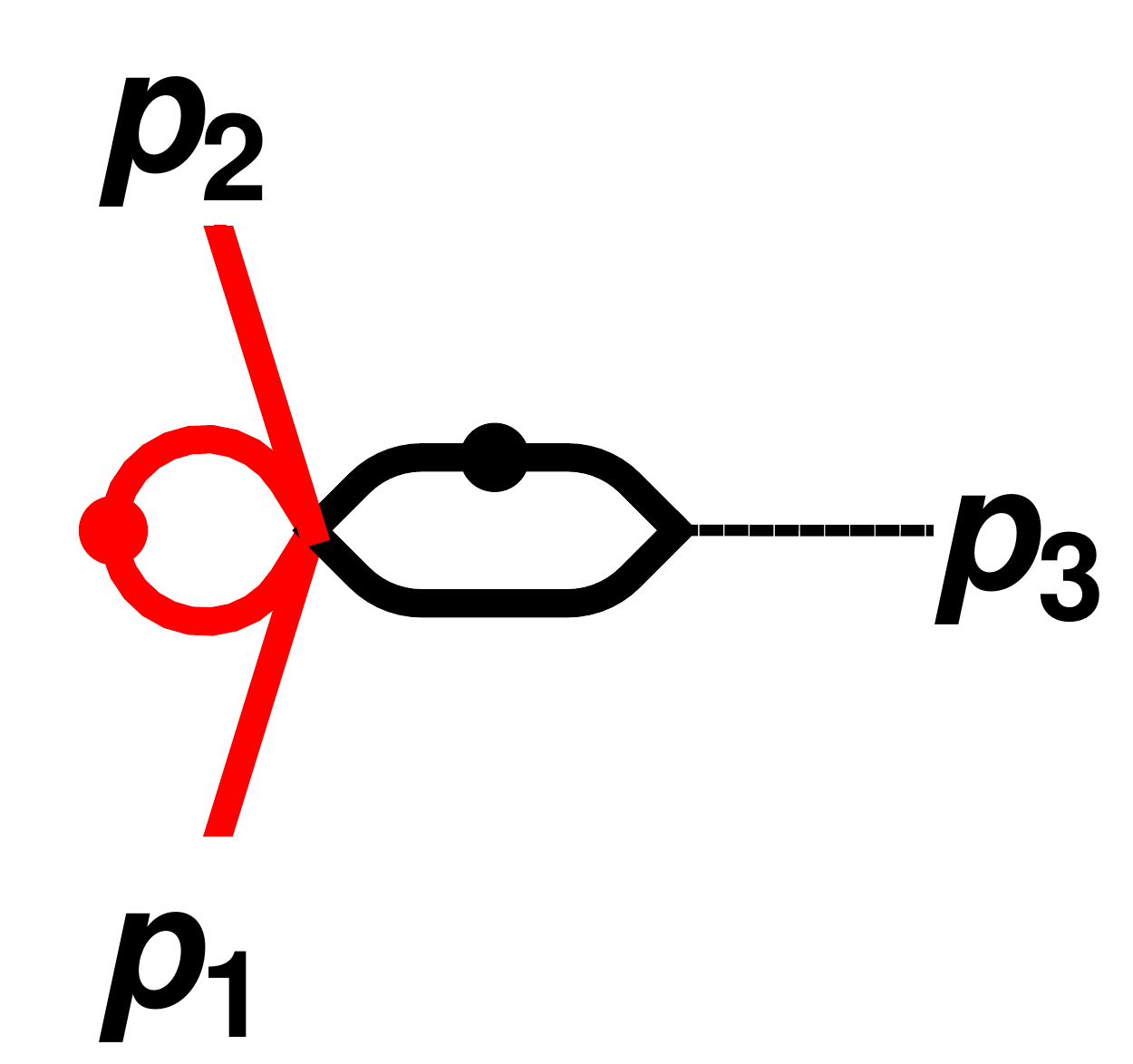}
  } \,
  \subfloat[$\mathcal{T}_9$]{%
    \includegraphics[width=0.105\textwidth]{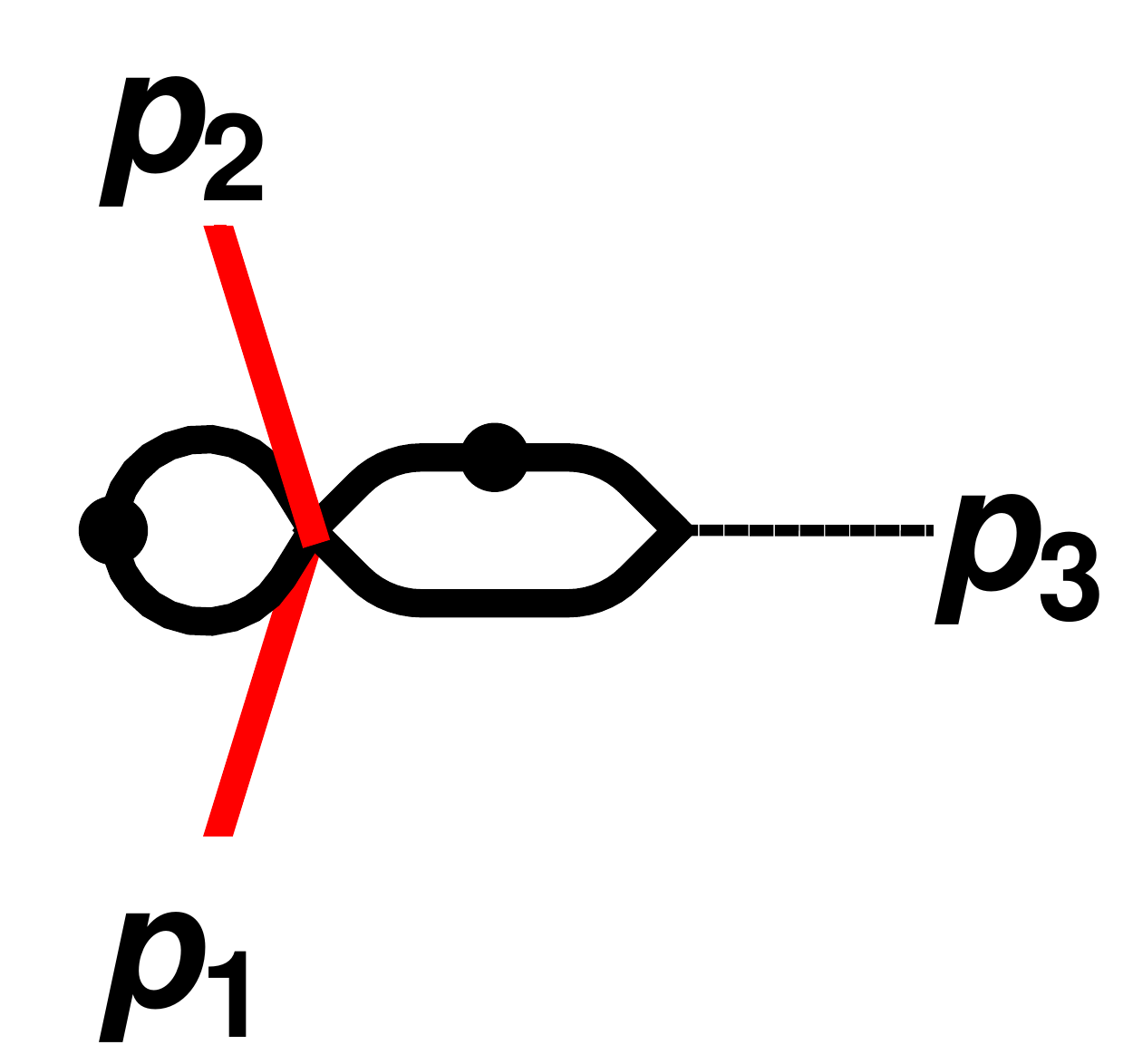}
  } \,
  \subfloat[$\mathcal{T}_{10}$]{%
    \includegraphics[width=0.105\textwidth]{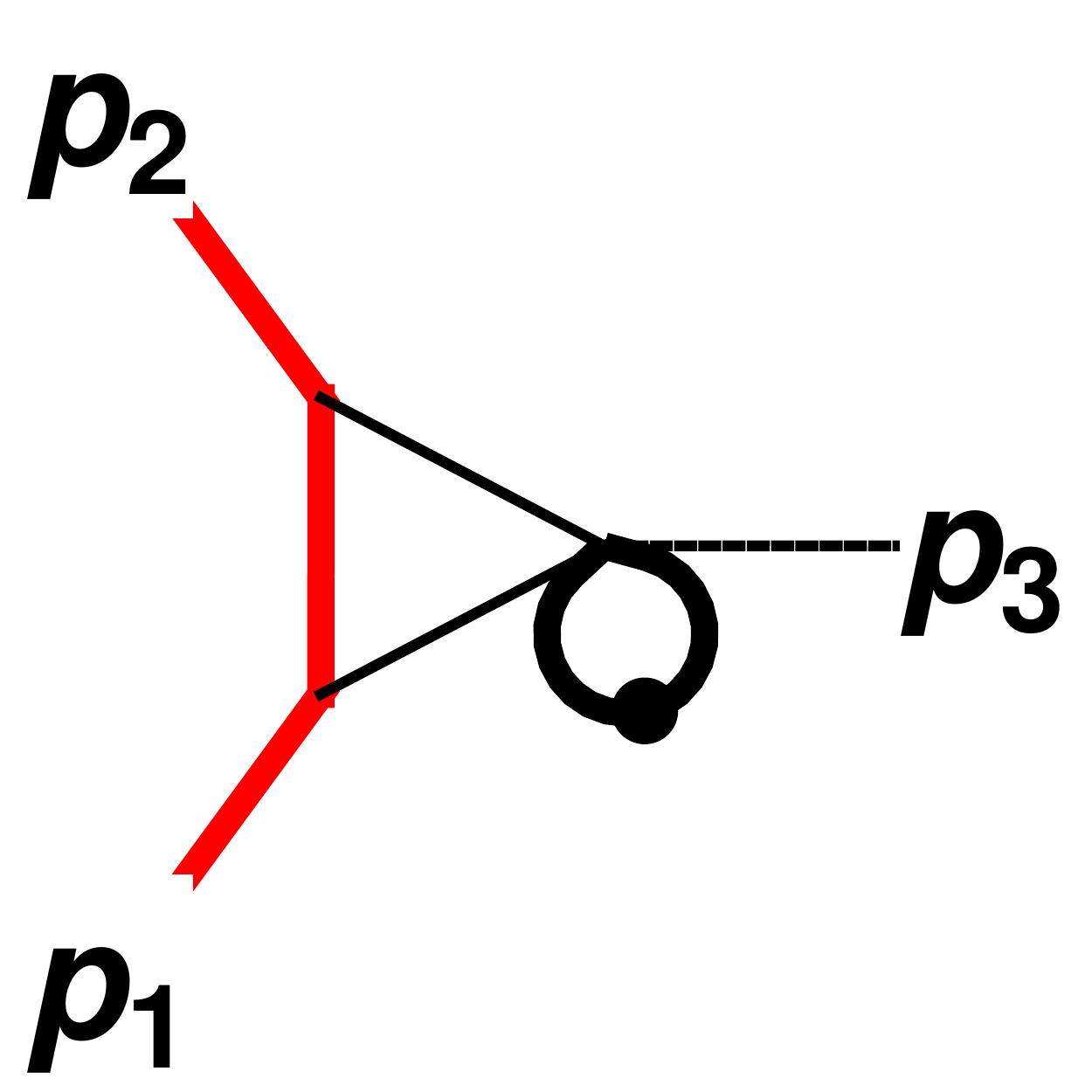}
  }\,
  \subfloat[$\mathcal{T}_{11}$]{%
    \includegraphics[width=0.105\textwidth]{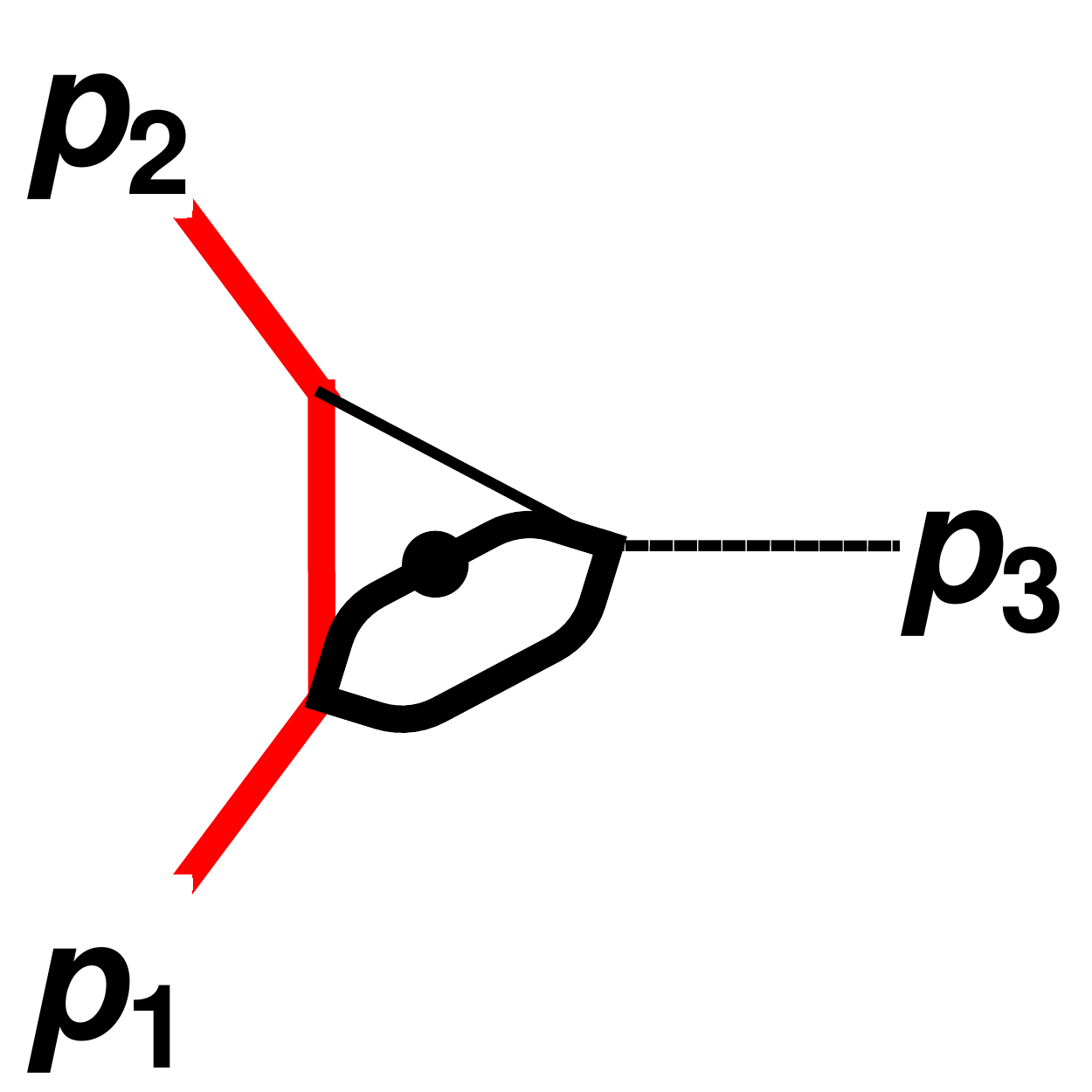}
  }\,
  \subfloat[$\mathcal{T}_{12}$]{%
    \includegraphics[width=0.105\textwidth]{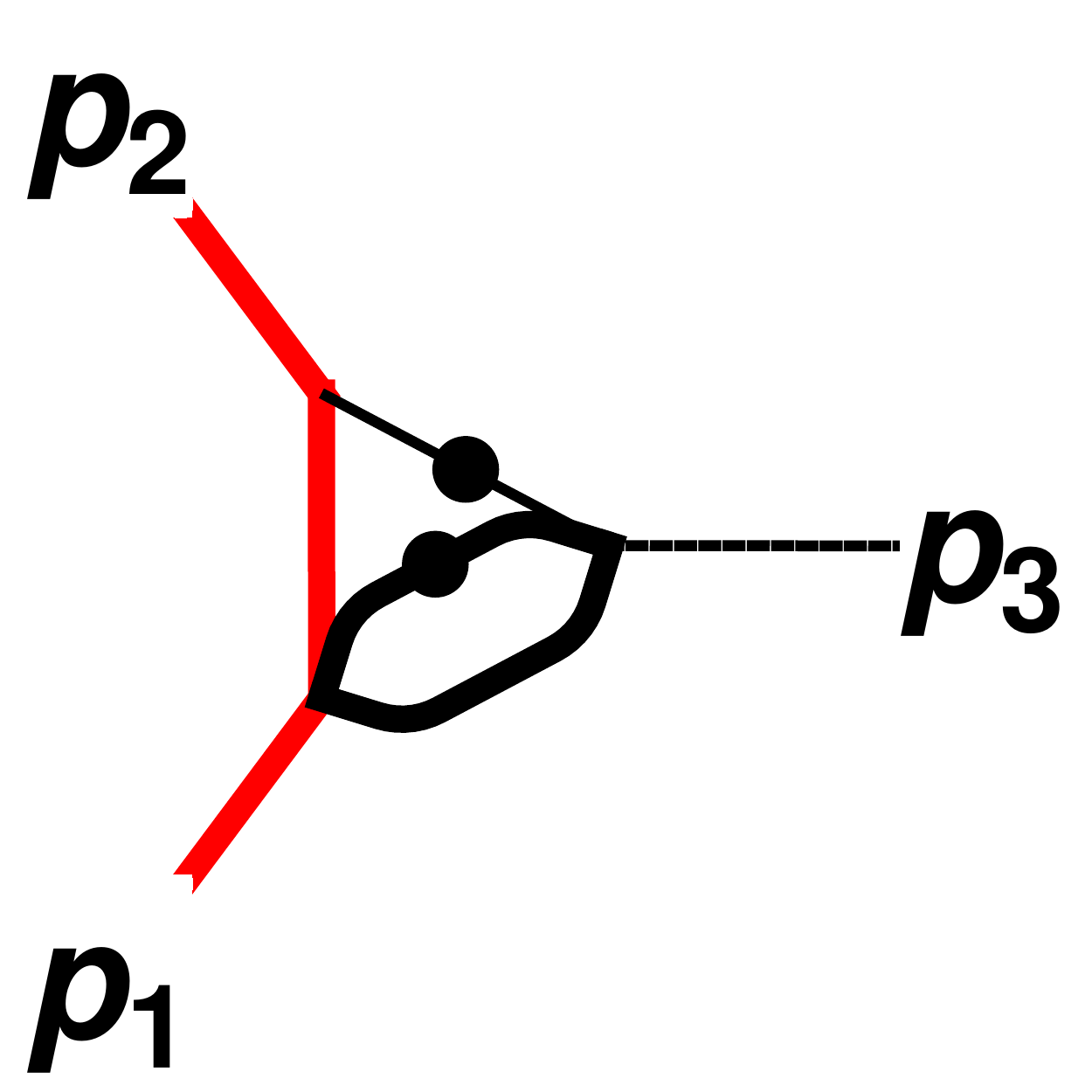}
  }\,
  \subfloat[$\mathcal{T}_{13}$]{%
    \includegraphics[width=0.105\textwidth]{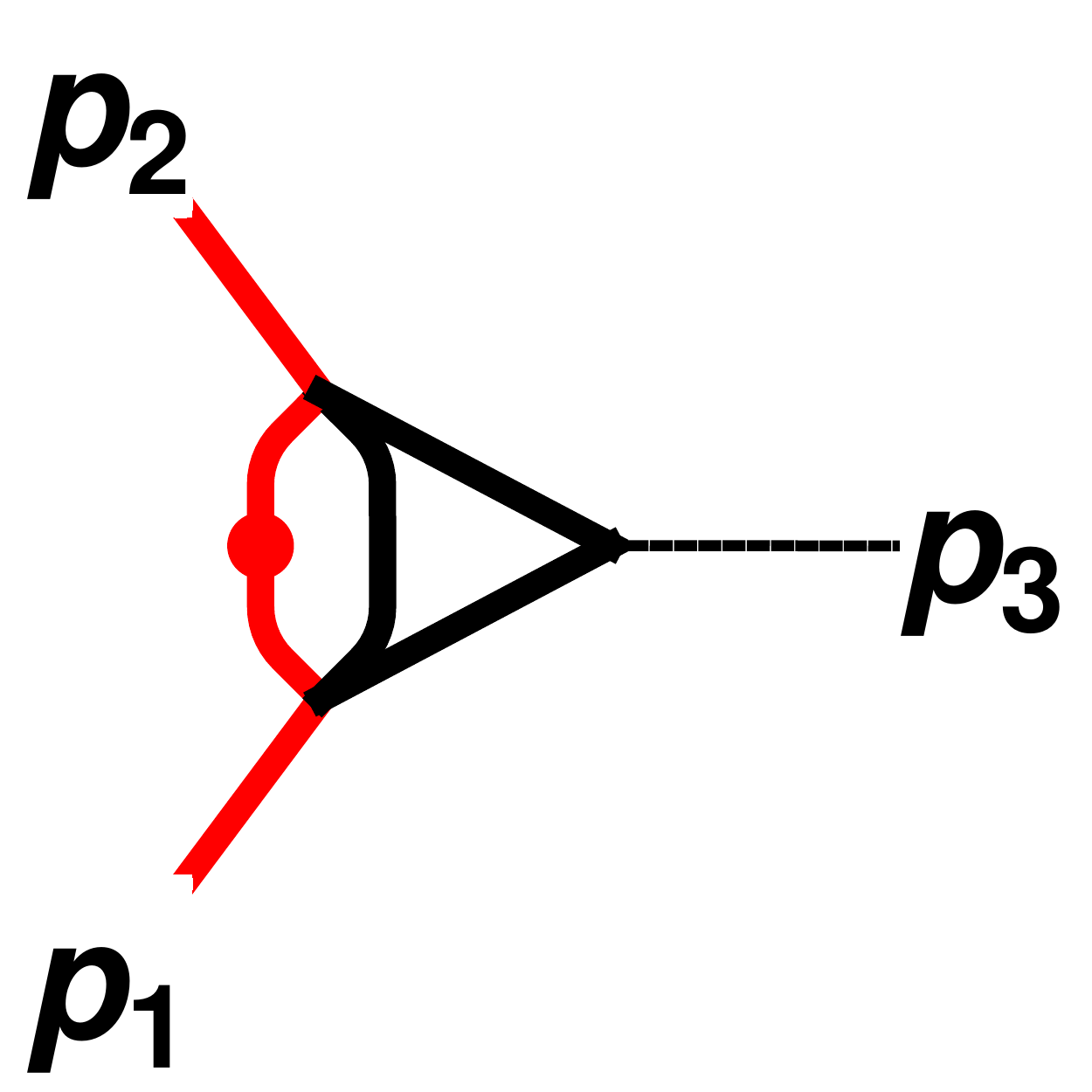}
  }\,
  \subfloat[$\mathcal{T}_{14}$]{%
    \includegraphics[width=0.105\textwidth]{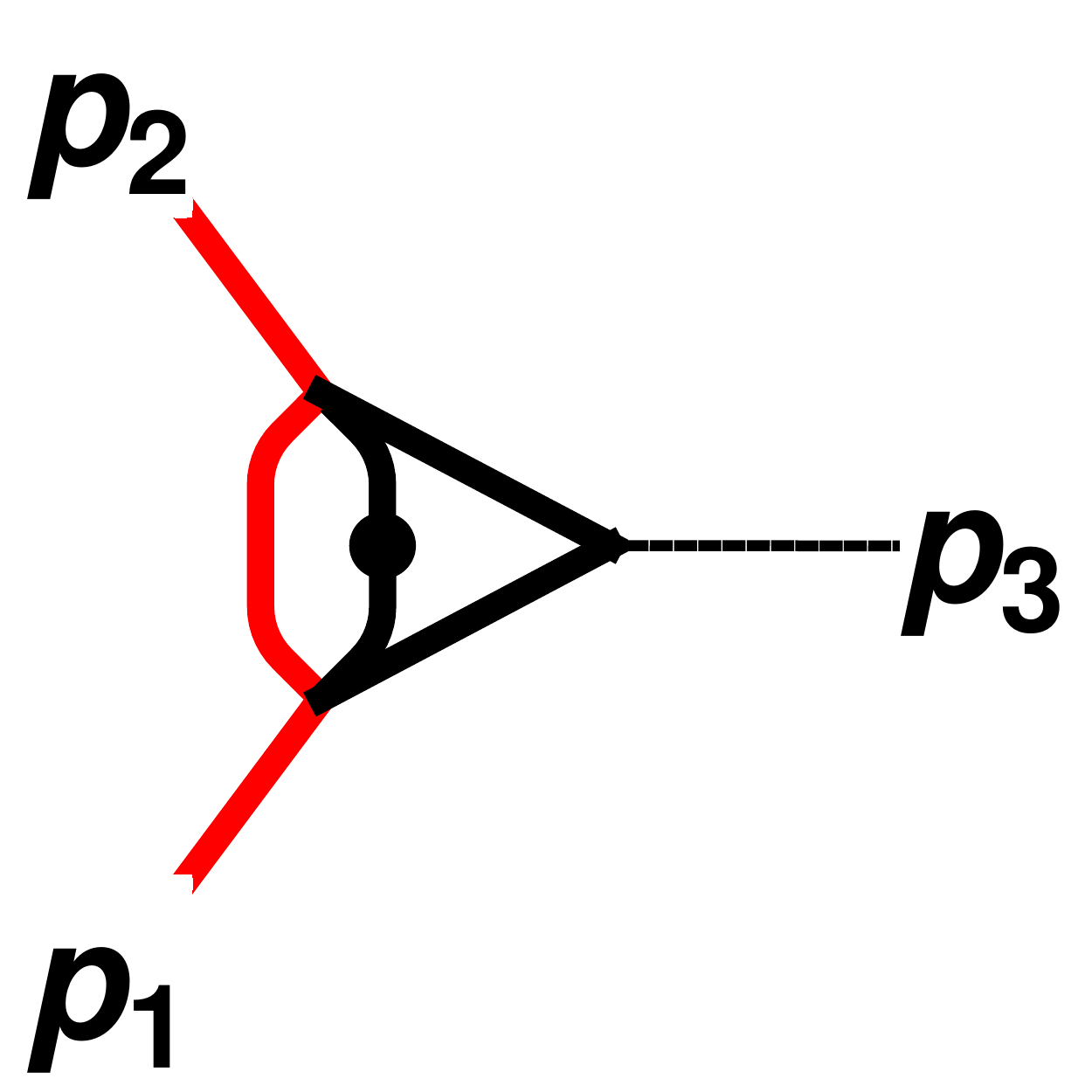}
  }\\
  \subfloat[$\mathcal{T}_{15}$]{%
    \includegraphics[width=0.105\textwidth]{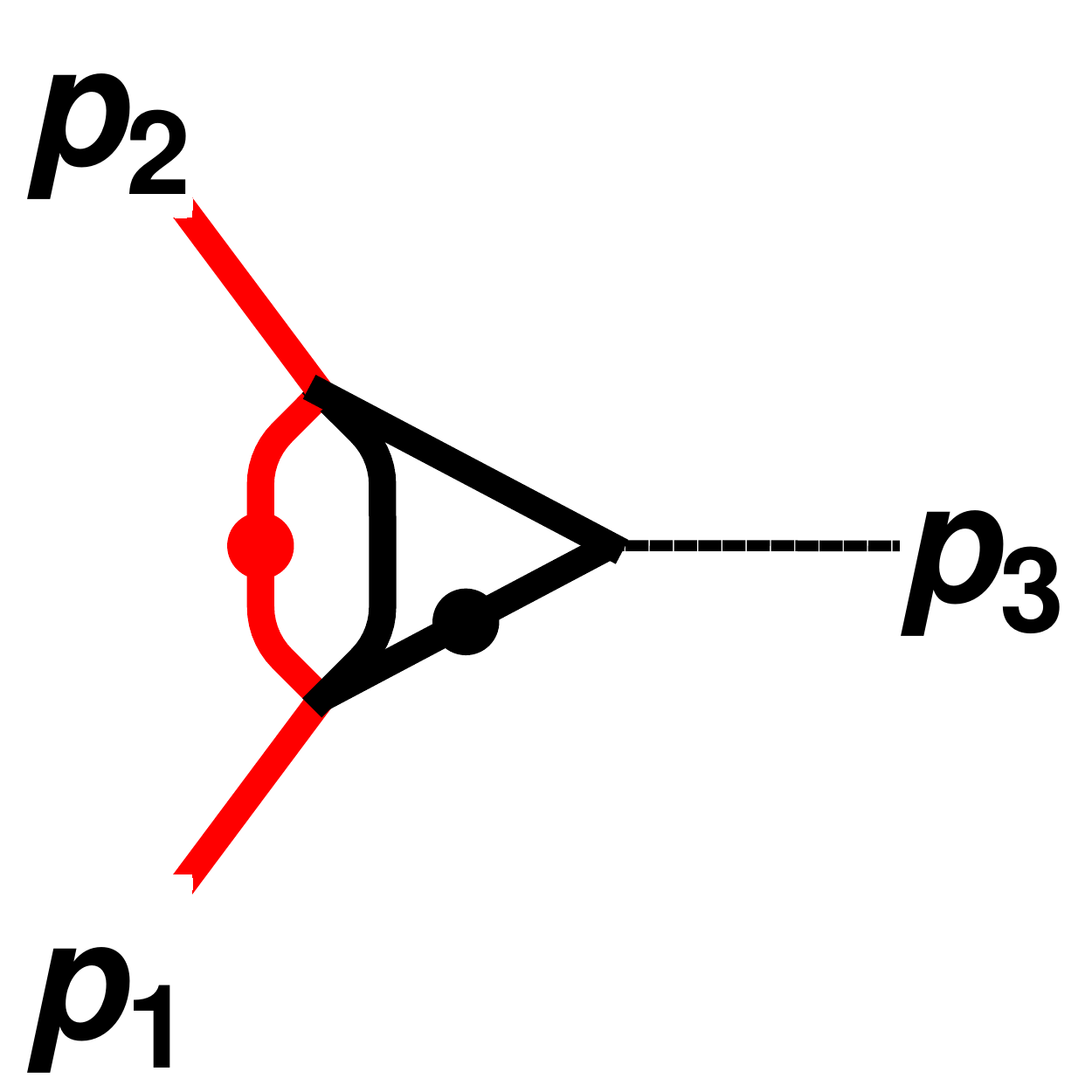}
  }\,
  \subfloat[$\mathcal{T}_{16}$]{%
    \includegraphics[width=0.105\textwidth]{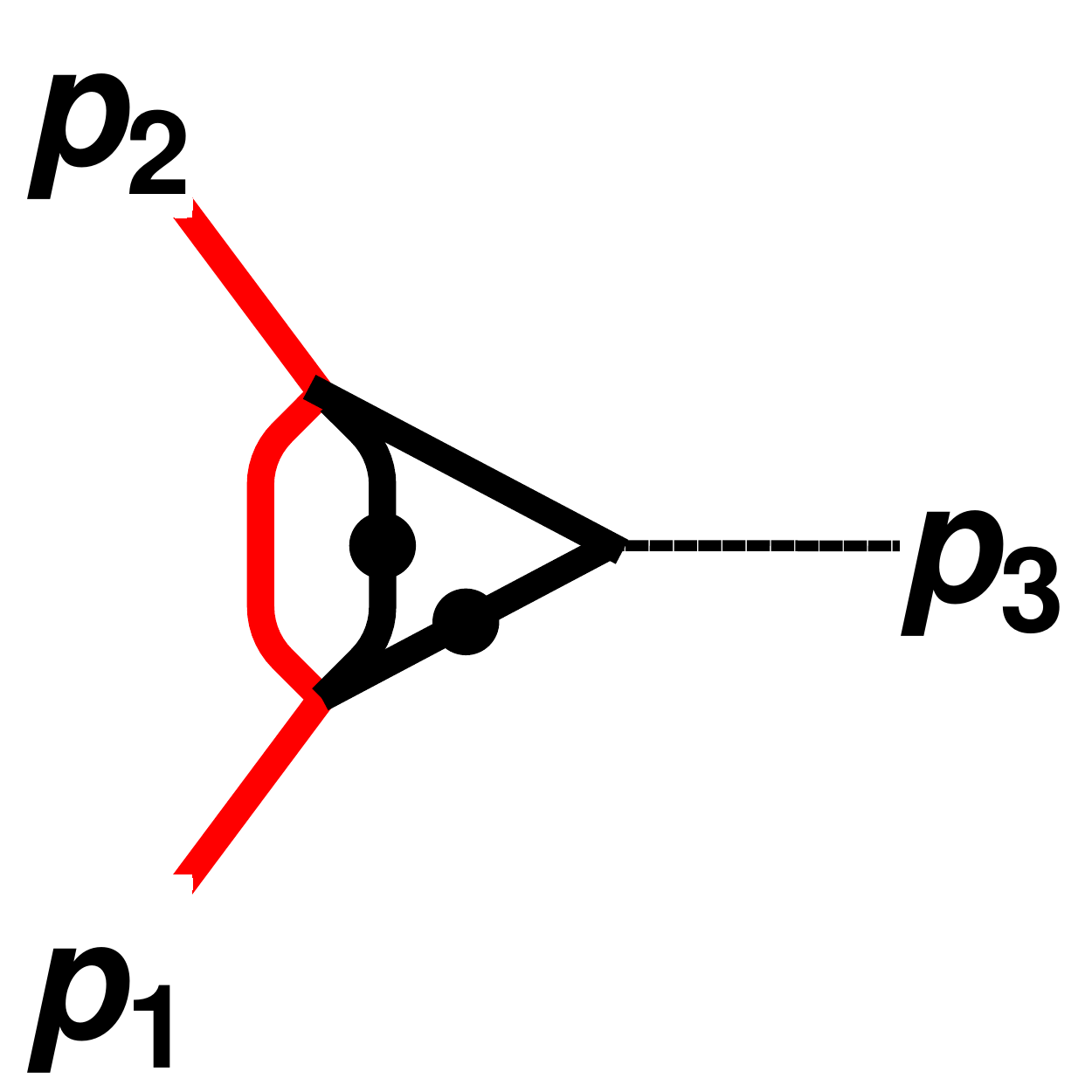}
  }\,
  \subfloat[$\mathcal{T}_{17}$]{%
    \includegraphics[width=0.105\textwidth]{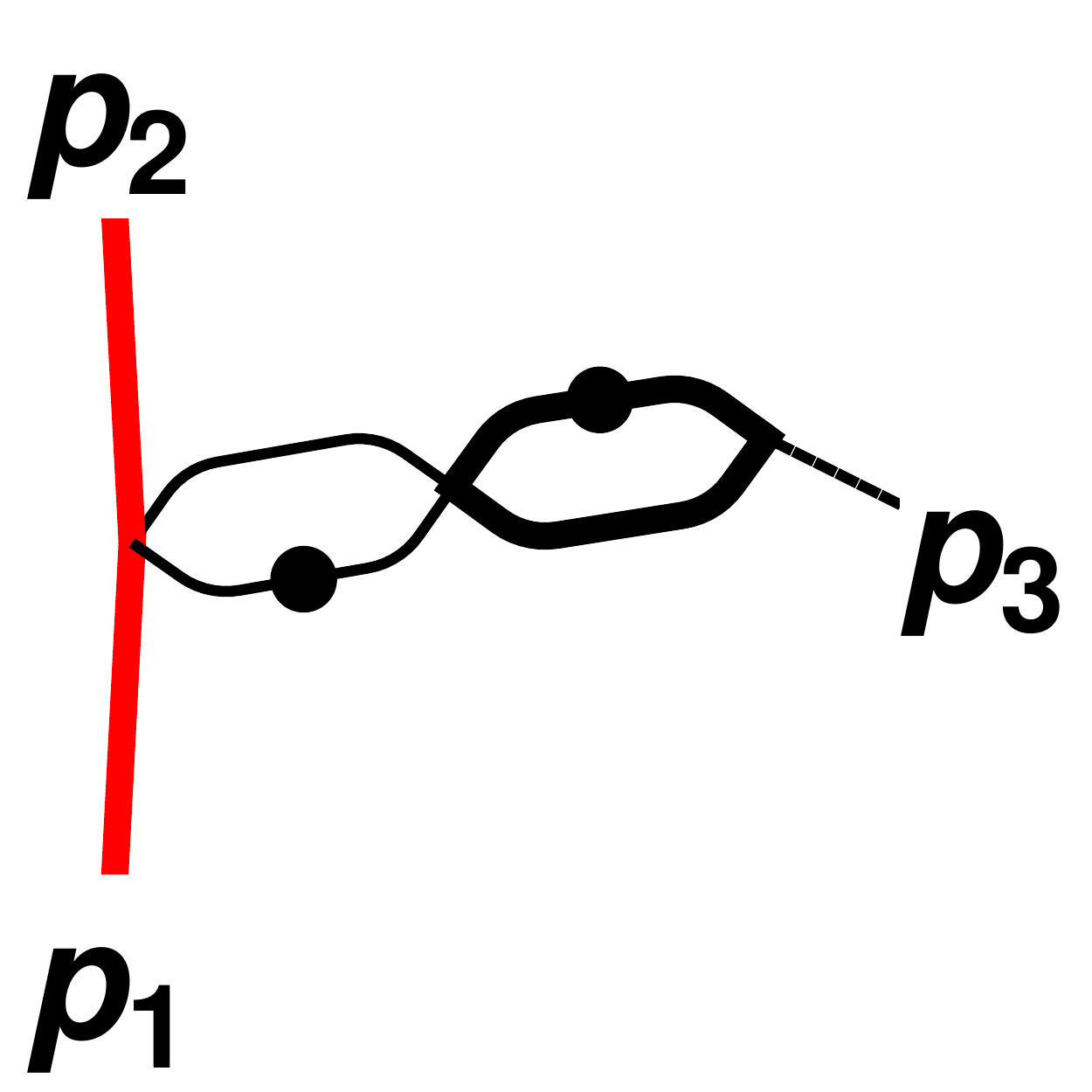}
  }\,
  \subfloat[$\mathcal{T}_{18}$]{%
    \includegraphics[width=0.105\textwidth]{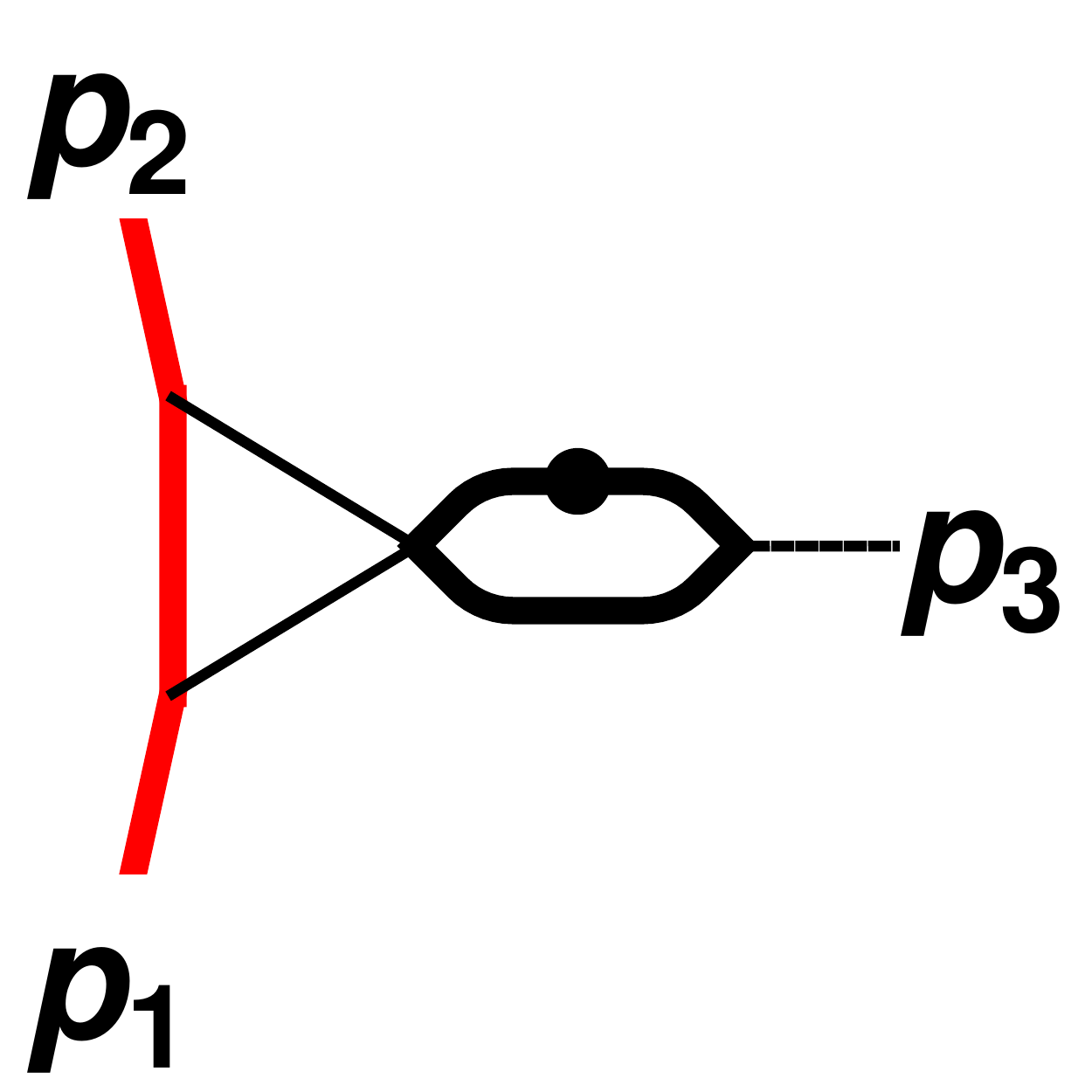}
  }\,
    \subfloat[$\mathcal{T}_{19}$]{%
    \includegraphics[width=0.105\textwidth]{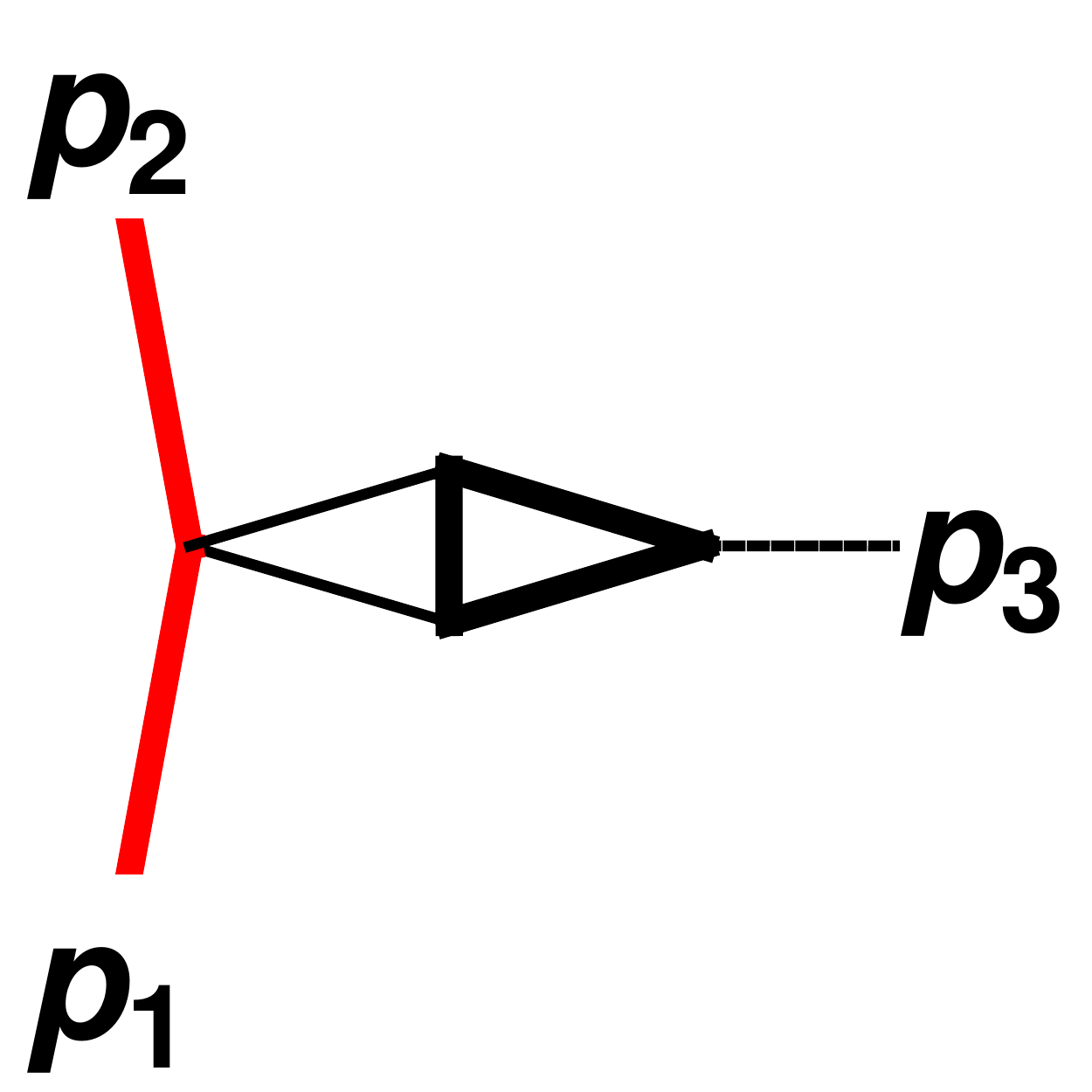}
  }\,
  \subfloat[$\mathcal{T}_{20}$]{%
    \includegraphics[width=0.105\textwidth]{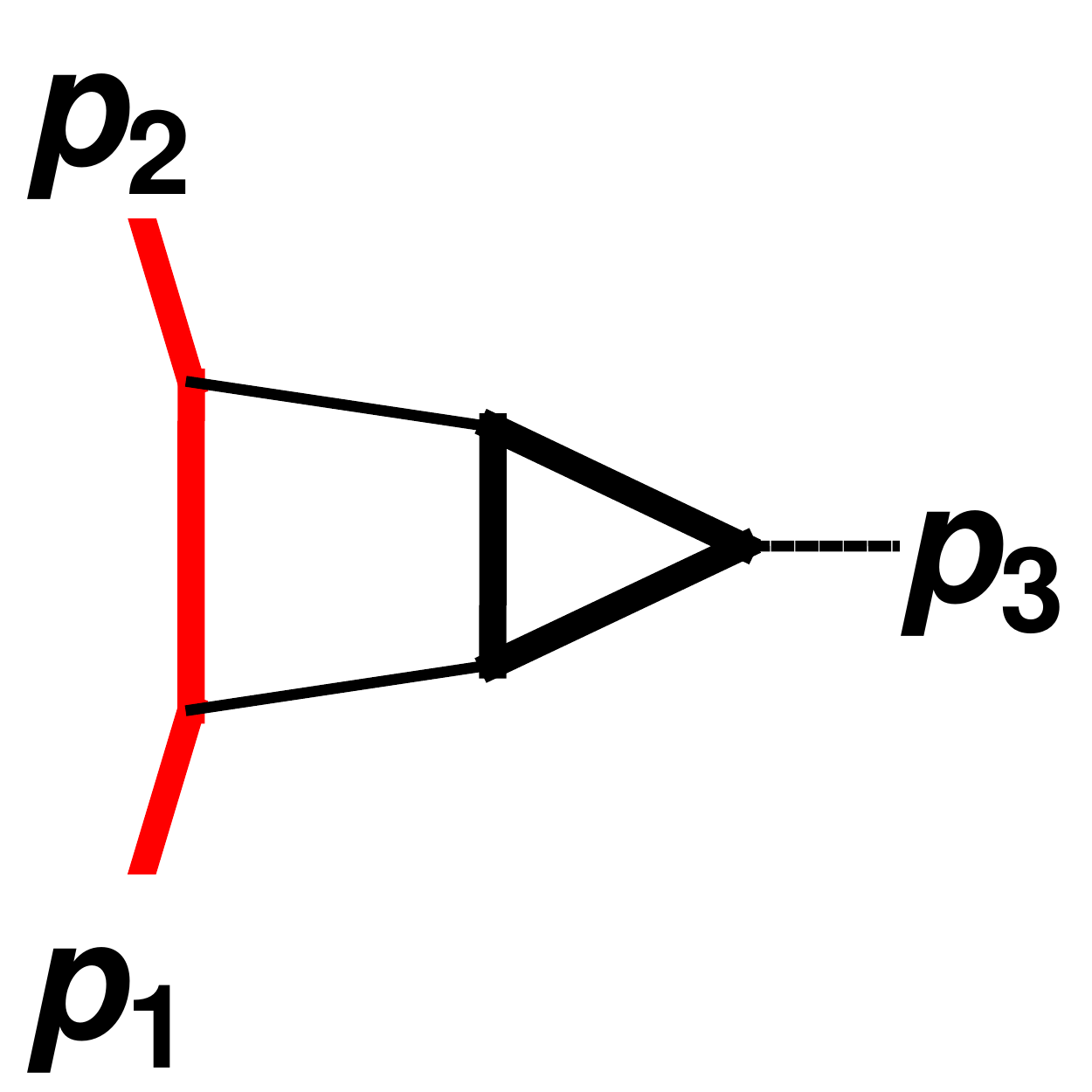}
  }
 \caption{Master integrals $\top{1\dots20}$ for the leading-order bottom Yukawa contributions to the Higgs boson decay into charm quarks. Thin black lines represent massless propagators, while thick black lines and red lines represent massive propagators with mass $m_b$ or $m_c$ respectively. Dots represent additional powers of the propagator. 
}
 \label{fig:MIs}
\end{figure}

In this appendix we provide the exact definitions of the master integrals satisfying the canonical differential equation~\eqref{eq:canonicalDEQ}:
\begin{align*}
  \GG_{1}&= \eps^2 \,\top{1}\,, &
  \GG_{2}&= \eps^2 \,  \top{2} \,, \nn
  \GG_{3}&= \eps^2 \, m_b^2 \, m_c^2   \, \top{3}\,, &
  \GG_{4}&=\eps^2 \, m_b^3  \, m_c \frac{\left(m_H^2-\lambda_{m_b} \right) \Lambda_{m_c} } {\left(m_H^2-\lambda_{m_c} \right) \Lambda_{m_b}} \left(2 \top{3}+\top{4} \right) \,, \nn
  \GG_{5}&= - \eps^2 \, m_b^2 \, m_H^2 \, \top{5}\,, &
  \GG_{6}&=- \eps^2 \, m_b^2 \, m_H^2   \, \top{6}\,,  \nn
  \GG_{7}&= -\eps^2 \, m_b^2 \, \lambda_{m_b} \left[ \left( \frac{1}{2} + \frac{m_H^2}{\lambda_{m_b}} \right) \top{6}+\top{7} \right] \,, &
  \GG_{8}&= - \eps^2 \, m_b^2 \, \lambda_{m_b} \, \top{8}\,, \nn
  \GG_{9}&= - \eps^2 \, m_b^2 \, \lambda_{m_b}  \, \top{9} \,,   &
  \GG_{10}&= \eps^3 \,  m_b^2 \, m_H^2 \frac{m_H^2-4m_c^2-\lambda_{m_c}}{m_H^2-\lambda_{m_c}} \, \top{10}\,, \nn 
  \GG_{11}&= \eps^3 \,  m_b^2 \, m_H^2 \frac{m_H^2-4m_c^2-\lambda_{m_c}}{m_H^2-\lambda_{m_c}}  \, \top{11}\,, & & \nn
  \GG_{12}&= \rlap{$ \displaystyle - \eps^2  \frac{ m_b^2 \, \lambda_{m_b}} {4 m_H^2} \left[ \left( m_H^2+\lambda_{m_c} \right)\left(\top{6}+2 \, \top{7} \right) + 4 \, m_b^2 \, m_c^2 \, \lambda_{m_c}\, \top{12} \right]$} && \nn 
  \GG_{13}&=\eps^3 \,  m_b^2 \, m_H^2 \frac{m_H^2-4m_c^2-\lambda_{m_c}}{m_H^2-\lambda_{m_c}}  \, \top{13}\,,  &
  \GG_{14}&=\eps^3 \,  m_b^2 \, m_H^2 \frac{m_H^2-4m_c^2-\lambda_{m_c}}{m_H^2-\lambda_{m_c}}  \, \top{14}\,, \nn
  \GG_{15}&=\rlap{$ \displaystyle  -\eps^2 \frac{m_b^2 \, \lambda_{m_b} \left( m_H^2+\lambda_{m_c} \right) }{4\,m_H^2} \bigg[ \top{6}+2\, \top{7} -4 \frac{m_b^2 \, m_c^2 \left(m_H^4-\lambda_{m_b}\lambda_{m_c}\right)}{\lambda_{m_b}} \top{12}    $} && \nn
  &\rlap{ $ \displaystyle  -4 \eps \frac{m_H^2}{\lambda_{m_b}} \left( 2\,\top{13} + \top{14} \right) \bigg]  - \eps^2(1+2\eps) \,m_b^4 \, m_c^2 \,m_H^2  \, \top{15}\,, $} && \nn
  \GG_{16}&= \eps^3 \, m_b^4 \, \lambda_{m_b}  \,\lambda_{m_c}    \,\top{16}  \,, & 
 \GG_{17}&= \eps^2 \, m_b^4 \, m_H^2 \lambda_{m_b}     \,\top{17}  \,, \nn 
  \GG_{18}&= \eps^3 \, m_b^4 \, \lambda_{m_b} \, \lambda_{m_c}    \, \top{18}\,, &
  \GG_{19}&= -\eps^3 \left(1-2\eps \right) m_b^2 \, m_H^2    \, \top{19} \,,  \nn
  \GG_{20}&= - \eps^4 \frac{m_b^4 \, m_H^4 \left( m_H^2-4m_c^2-\lambda_{m_c}\right)}{m_H^2-\lambda_{m_c}}   \, \top{20}\,, &
\label{def:CanBasis}
\stepcounter{equation}\tag{\theequation}
\end{align*}
where we defined the abbreviations $\lambda_{m}=\sqrt{m_H^2}\,\sqrt{m_H^2-4\,m^2}$ and $\Lambda_m=\sqrt{m_H^2-2\,m^2-\lambda_{m}}$. The integrals $\top{i}$ appearing above are depicted in fig.~\ref{fig:MIs}. The definitions and expressions of the canonical master integrals are also given in the ancillary files accompanying the \texttt{arXiv} submission of this publication.

\bibliographystyle{JHEP}

\bibliography{Hbb}

\end{document}